\documentclass[acmsmall,screen]{acmart}

\usepackage{url}
\usepackage{xspace}
\usepackage{wrapfig}
\usepackage{hyperxmp}
\usepackage{booktabs}
\usepackage{makecell}
\usepackage{hyperref}
\usepackage{graphicx}
\usepackage{colortbl}
\usepackage{tabularx}
\usepackage{multirow}
\usepackage{enumitem}
\usepackage{subfigure}
\usepackage{todonotes}
\usepackage{tablefootnote}
\usepackage{color, xcolor}
\usepackage{amsthm, amsmath, amsfonts}
\newcolumntype{L}[1]{>{\raggedright\arraybackslash}p{#1}}
\newcolumntype{C}[1]{>{\centering\arraybackslash}p{#1}}
\newcolumntype{Y}{>{\raggedright\arraybackslash}X}

\newcommand{\eg}{\textit{e.g.,}\xspace}

\newcommand{\figref}[1]{Fig.~\ref{#1}\xspace}
\newcommand{\tabref}[1]{Table~\ref{#1}\xspace}
\newcommand{\secref}[1]{Section~\ref{#1}\xspace}
\newcommand{\appref}[1]{Appendix~\ref{#1}\xspace}

\newsavebox{\queryboxcontent}
\newenvironment{querybox}{%
  \par\smallskip
  \begin{lrbox}{\queryboxcontent}%
  \begin{minipage}{0.94\columnwidth}%
}{%
  \end{minipage}%
  \end{lrbox}%
  \noindent\makebox[\columnwidth][c]{%
    \setlength{\fboxsep}{6pt}%
    \setlength{\fboxrule}{0.5pt}%
    \fbox{\usebox{\queryboxcontent}}%
  }%
  \par\smallskip
}

\setcopyright{acmlicensed}
\copyrightyear{2026}
\acmYear{2026}
\acmDOI{XXX.XXX}

\acmJournal{CSUR}
\acmVolume{1}
\acmNumber{1}
\acmArticle{1}
\acmMonth{1}

\begin{document}

\title{Toward Secure LLM Agents: Threat Surfaces, Attacks, Defenses, and Evaluation}

\author{Yuchen Ling}
\affiliation{\institution{State Key Laboratory for Novel Software Technology, Nanjing University}\city{Nanjing}\country{China}\postcode{210093}}
\email{yuchenling@smail.nju.edu.cn}
\orcid{0009-0006-9227-3824}

\author{Shengcheng Yu}
\affiliation{\department{School of CIT, MDSI, IAS}\institution{Technical University of Munich}\city{Heilbronn}\country{Germany}\postcode{74076}}
\email{shengcheng.yu@tum.de}
\orcid{0000-0003-4640-8637}
\authornote{Shengcheng Yu and Chunrong Fang are the corresponding authors.}

\author{Zhenyu Chen}
\affiliation{\institution{State Key Laboratory for Novel Software Technology, Nanjing University}\city{Nanjing}\country{China}\postcode{210093}}
\email{zychen@nju.edu.cn}
\orcid{0000-0002-9592-7022}

\author{Chunrong Fang}
\affiliation{\institution{State Key Laboratory for Novel Software Technology, Nanjing University}\city{Nanjing}\country{China}\postcode{210093}}
\authornotemark[1]
\email{fangchunrong@nju.edu.cn}
\orcid{0000-0002-9930-7111}

\begin{abstract}
Large language model (LLM) agents have recently attracted increasing attention and become an important research direction in software engineering.
They are rapidly moving from conversational interfaces to software components that plan, invoke tools, maintain memory, and act on external environments in agentic loops.
As LLMs become embedded in these action-oriented loops, the security problem changes as well. In LLM agents, failures are no longer limited to unsafe text generation, since untrusted content may redirect control flow, misuse tool privileges, corrupt persistent state, leak sensitive information, or trigger harmful external actions.
These risks depend on how probabilistic models are integrated with orchestration logic, tools, state stores, permission mechanisms, and runtime controls.
Research on these risks is expanding quickly, but it remains fragmented across attack families, defense layers, application domains, and evaluation settings.
In this survey, we synthesize 247 papers through a lifecycle-based, systems-oriented framework, modeling LLM agent security around the interaction of information flow, delegated authority, and persistent state. We organize the literature around four research questions that ask how LLM agent security should be modeled, which threat surfaces and attack families dominate, what defenses have been proposed and with what tradeoffs, and how security claims are evaluated. We find that prompt injection and tool-mediated control-flow hijacking still dominate the field, while persistent state corruption and multi-agent propagation are becoming central emerging concerns. We further find that current defenses provide useful building blocks but remain weakly compositional, and that existing benchmarks still underrepresent long-horizon, stateful, and deployment-sensitive risks.
From a software engineering perspective, we argue that secure LLM agents require explicit trust boundaries, principled privilege control, provenance-aware state management, and evaluation practices that
support architecture design, testing, runtime assurance, and deployment under realistic operational conditions.
\end{abstract}



\maketitle

\section{Introduction}

Large language model (LLM) agents are increasingly serving as decision-making and control components in intelligent software systems \cite{liu2024personal, yu2025survey}. Unlike text-only LLM applications, these systems can decompose tasks, invoke tools, browse the web, manipulate files, execute code, maintain memory, and coordinate with other agents \cite{openai2025chatgpt, anthropic2024developing, liu2025towards}. Each added capability gives the supporting LLM a larger role in shaping what the surrounding software does. As a result, the security problem expands from content generation to operational behavior. Once an LLM is embedded in agentic loops that connect language understanding, planning, action selection, tool execution, and state update, failures may appear as hijacked workflows, unauthorized tool use, persistent state corruption, sensitive data leakage, or harmful external actions rather than as problematic text alone \cite{zhan2024injecagent, li2024agentpoison, shmatikov2025multi}. In agentic settings, a malicious instruction hidden in a web page, a poisoned tool response, or a contaminated memory entry can affect not only the next LLM agent output, but also later decisions, stored state, and downstream LLM agents \cite{xiao2024wipi, dong2025memory, tiwari2024prompt}. Existing overviews likewise emphasize that agentic systems combine autonomy, tool use, and deployment risk in ways that go beyond prompt-only LLM safety \cite{ji2024navigating, deng2025ai, he2026emerged}. LLM agent security should therefore be understood as a software and systems security problem.

This shift is consequential because LLM agents operate under delegated authority. They not only interpret language but also choose actions, invoke capabilities, and modify persistent state on behalf of users or organizations. As a result, the central security question is no longer only whether a model can be induced to say something unsafe, but whether untrusted or invalid content can alter control flow, cross privilege boundaries, compromise state integrity, or amplify across coordinated workflows. The resulting behaviour is not determined by the model alone. It emerges from the interaction among LLM outputs, orchestration code, tool interfaces, state stores, permission configurations, monitoring components, and human oversight.

Unlike conventional software systems, whose control decisions are largely encoded in explicit program logic, LLM agents or LLM-agent-based systems delegate part of goal interpretation, planning, and action selection to a probabilistic model. Similar or even the same inputs may therefore lead to different trajectories, while changes in context composition or model versions may alter behaviour even when the surrounding code remains unchanged. Security cannot be established only through deterministic input-output expectations or one-time validation. It also requires methods that characterize behavioural variation, exercise long-horizon interactions, and provide runtime evidence when uncertain decisions cross authority boundaries. LLM agent security therefore sits at the intersection of artificial intelligence, systems security, and software engineering because model behavior, architecture, interface contracts, testing strategies, observability, deployment configuration, and system evolution jointly determine whether model-level failures become operational incidents. This perspective brings LLM agent security closer to classical concerns in systems security, including trust boundaries, mediation, capability control, provenance, and containment, while extending them to software systems whose control decisions are partly expressed through natural-language reasoning and flexible context composition.

Despite rapidly growing attention, the literature on LLM agent security remains fragmented, uneven in emphasis, and incomplete in its treatment of deployment risk. Recent work has examined prompt injection, tool-use security, memory poisoning, runtime monitoring, secure agent architectures, and multi-agent coordination \cite{zhan2024injecagent, ye2024toolsword, li2024agentpoison, li2024guardagent, wang2024netsafe}. Adjacent surveys, however, frame the space from broader trustworthy-agent, governance, or full-stack safety perspectives rather than from one shared agentic-loop model \cite{deng2025ai, yu2025survey, liu2025comprehensive, raza2026trism}. The result is a field rich in individual attacks, defenses, and benchmarks, but still lacking a coherent view of how these pieces fit together. The imbalance of attention is also notable. Prompt injection has become the most visible topic and has driven substantial progress, yet persistent state integrity, tool-mediated privilege misuse, and multi-agent propagation are increasingly important in realistic deployments while remaining less systematically synthesized \cite{li2024agentpoison, wang2025model, shin2025seven, arremsetty2026security}. Meanwhile, current benchmark practice still emphasizes immediate attack success in bounded environments, leaving long-horizon, stateful, and deployment-sensitive risks insufficiently evaluated \cite{tramer2024agentdojo, zhang2025agent, yuan2024r, chen2024safeagentbench}.

In this survey, we address this gap through a curated corpus of 247 papers and a lifecycle-based, systems-oriented analytical framework. We study how risks emerge and propagate across inputs, planning, decisions, tool execution, outputs, memory, monitoring, and coordination in existing research. This structure allows us to connect prompt injection, memory corruption, tool-mediated abuse, and multi-agent risk within a shared account of the agentic loop, and to compare defenses and evaluations according to where they intervene and what assumptions they make. Existing surveys provide useful background, but they do not yet offer the same combination of lifecycle structure, systems interpretation, deployment and assurance implications, and transparent cross-sectional annotation that we aim to provide \cite{he2025security, zhu2025survey, minnich2025systematization}.

We organize the survey around four research questions that ask how LLM agent security should be modeled, which threat surfaces and attack families dominate current research, what defense mechanisms have been proposed and with what tradeoffs, and how security claims are evaluated. This design lets us move beyond descriptive aggregation towards conceptual synthesis while also capturing how quickly the field is changing. In our corpus, the literature grows from 3 papers in 2023 to 42 in 2024, 121 in 2025, and 81 papers collected by April 27, 2026. The pace of growth reflects strong momentum, but it also signals that terminology, methodological norms, and deployment assumptions are still unsettled. A structured synthesis is therefore necessary not only to summarize the field, but also to clarify its emerging boundaries, dominant patterns, and unresolved gaps.

The main contributions of this survey are as follows.
\begin{itemize}[leftmargin=6mm]
\item We construct and analyze a curated corpus of 247 papers on LLM agent security through a hybrid review pipeline that combines database retrieval, LLM-assisted candidate expansion with web search, citation snowballing, and manual screening.
\item We propose a lifecycle-based and systems-oriented framework that connects inputs, planning, decisions, tool execution, outputs, memory, monitoring, and multi-agent coordination, providing a basis for reasoning about LLM agent security from a software engineering perspective. 
\item We synthesize the threat landscape of LLM agents, showing how prompt injection, tool-mediated control-flow hijacking, persistent state corruption, and multi-agent propagation form recurring and interrelated risk families.
\item We compare existing defense strategies by their intervention points, trust assumptions, utility costs, and composability limits, and we analyze the benchmark ecosystem in terms of coverage, metrics, methodological gaps, and reporting gaps.
\end{itemize}

The remainder of this paper is organized as follows. \secref{sec:background} introduces the technical background and security motivation of LLM agents. \secref{sec:method} and \secref{sec:trend} present the survey methodology and corpus-level observations. \secref{sec:rq1} to \secref{sec:rq4} answer the four research questions. \secref{sec:discussion} discusses broader implications and research priorities, \secref{sec:validity} summarizes the main threats to validity, and \secref{sec:conclusion} concludes the paper. To support broader dissemination and easier exploration of the curated corpus, we also provide a \href{https://ling-yuchen.github.io/LLMAgentSecuritySurvey/}{website} for this survey.

\section{Background}\label{sec:background}

Understanding LLM agent security requires clarifying what distinguishes agents from ordinary LLM applications. Language models are increasingly embedded in systems that plan, invoke tools, update state, and affect external environments. Their security properties therefore depend on both the capabilities that enable autonomy and the risks created when language-mediated reasoning is connected to operational authority.

\subsection{Rise of LLM-based Autonomous Agents}\label{sec:b1}

Recent advances in large language models (LLMs) have shifted research attention from passive text generation towards action-oriented systems that can reason, use tools, access external environments, and pursue multi-step goals \cite{liu2024personal, han2025survey, zhu2025survey}. Early work on reasoning-acting integration and tool use already showed that LLMs can serve not only as conversational interfaces but also as decision-making components in interactive software systems \cite{yao2023react, schick2023toolformer}. That shift has since accelerated through the emergence of web agents, coding agents, memory-augmented assistants, embodied agents, and multi-agent workflows, where LLMs are used to interpret instructions, plan actions, execute tasks, revise intermediate results, and coordinate subtasks \cite{zhou2023webarena, yang2024sweagent, packer2023memgpt, wang2023voyager, wu2023autogen}.

A defining feature of contemporary LLM agents is that they function as autonomous or semi-autonomous workflow components rather than as answer generators alone. In research settings, agents can browse websites \cite{zhou2023webarena, kang2024llm}, manipulate files \cite{openai2025operator, anthropic2024developing}, generate and execute code \cite{yang2024sweagent, kang2024llmb}, decompose and coordinate tasks \cite{wu2023autogen, wang2024netsafe, kang2026teams}, and maintain persistent state across long-horizon interactions \cite{packer2023memgpt, li2024agentpoison, dong2025memory}. Comparable capabilities are increasingly visible in deployed systems, including browser-use agents \cite{openai2025operator, deepmind2025gemini}, computer-use tools \cite{anthropic2024developing, anthropic2025claude}, and agentic assistants with remote browsing, terminal access, and connector-based access to external applications \cite{openai2025chatgpt, openai2026gpt}. The convergence between research prototypes and deployment-oriented systems makes the security implications of these capabilities increasingly difficult to treat as speculative.

This development also changes the practical meaning of autonomy. An LLM agent is not merely producing a response to be read by a user. It may select the next operation in a workflow, decide which page to open, choose which tool to invoke, determine what command to run, update memory, or communicate with other agents. As the operational reach of these systems expands, their failure modes increasingly resemble those of software systems with delegated authority. The relevant question is no longer only whether the model outputs incorrect or unsafe text, but whether it makes unsafe decisions, takes unsafe actions, or preserves unsafe state while acting on behalf of end users or organizations.
The rising LLM agents and LLM-agent-based systems are different from traditional software systems whose workflows are basically deterministically encoded.
The distinctive issue in LLM-agent-based systems is that a probabilistic model participates directly in interpreting context and selecting control transitions. Agent behaviour can vary with sampling, prompt composition, retrieved content, accumulated state, and model updates, making some security-relevant decisions difficult to specify as fixed mappings from inputs to outputs. This uncertainty is not confined to the model boundary once generated plans and tool calls affect persistent state or external systems. It becomes a property of the larger software system and complicates reproducibility, testing, and assurance across repeated and evolving deployments.

\subsection{Security Challenges and Deployment Risks of LLM Agents}\label{sec:b2}

LLM agent security should be understood as a systems problem in which uncertainty becomes consequential through action. It arises when language-mediated reasoning is coupled with action selection, state management, and externally consequential execution. Unlike content-level safety, which is concerned with prompts and responses, agent security concerns how user instructions, retrieved content, tool outputs, memory updates, and inter-agent messages can alter control flow, misuse privileges, corrupt state, or trigger harmful operations.

From this perspective, an LLM agent is defined less by its application domain than by an agentic loop in which the LLM participates in goal interpretation, planning, tool selection, action execution, state update, progress monitoring, or coordination with other agents \cite{song2026framework, ji2026agentic, raza2026trism}. This framing allows web agents, coding agents, memory-augmented assistants, embodied systems, and multi-agent workflows to be analyzed within a shared security vocabulary \cite{arremsetty2026security, shin2025seven, liu2026thinker}. It also clarifies why LLM agent security cannot be reduced to generic LLM security. In text-only settings, unsafe behavior typically produces problematic textual content \cite{liu2025comprehensive, ma2025safety}. In agentic settings, the same underlying vulnerability may instead redirect a workflow, invoke privileged tools, leak sensitive data, poison persistent or temporary memory, or propagate malicious instructions to downstream agents. As a result, securing LLM agents requires attention to control-flow integrity, authority boundaries, runtime containment, state provenance, and monitoring of external actions \cite{iqbal2025isolategpt, song2025progent, beutel2024instruction, kiciman2024defending}.

These risks are already visible in both research prototypes and deployed systems. Prior studies show that agents can be redirected by malicious web content \cite{xiao2024wipi, wang2025webinject}, induced to exfiltrate data through tools \cite{yair2025invitation, he2025unveiling}, manipulated into unsafe software actions \cite{xie2025red, lo2025your}, or compromised through poisoned memory and inter-agent communication \cite{li2024agentpoison, tiwari2024prompt, shmatikov2025multi}. Deployment documents for browser- and computer-using agents similarly identify prompt injection, harmful task execution, model error, irreversible actions, and risks created by remote browsing, terminal access, and connector-based data access as major concerns \cite{openai2025operator, openai2025chatgpt, anthropic2024developing, deepmind2025gemini, openai2026gpt, anthropic2025claude}. These examples matter because failures occur at the boundary where language is translated into operational authority. The central challenge is therefore to secure an agentic loop that turns natural language into decisions and decisions into external effects.

\section{Survey Methodology}\label{sec:method}

\begin{figure}[!htbp]
\centering
\includegraphics[width=0.9\textwidth]{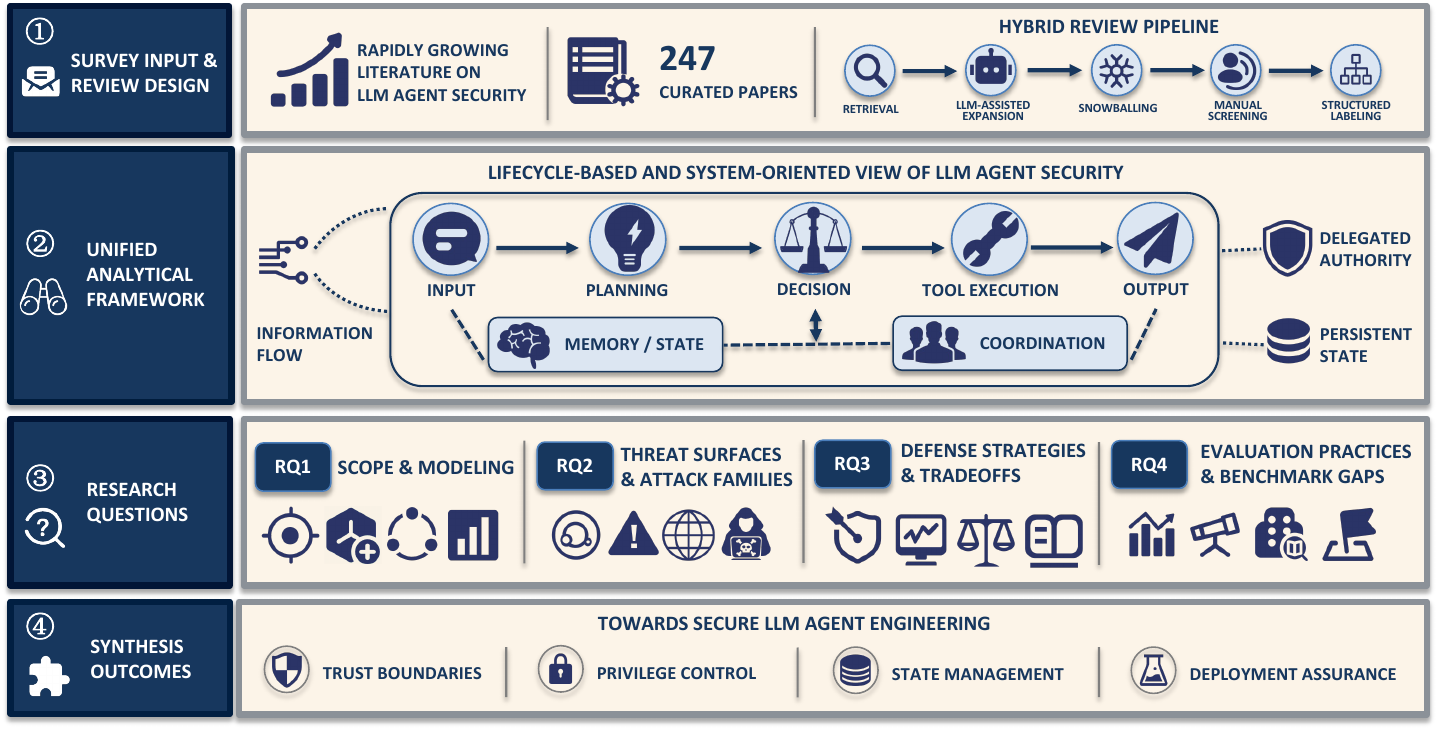}
\caption{Analytical Framework of This Survey on LLM Agent Security}
\label{fig:survey-framework}
\end{figure}

\figref{fig:survey-framework} provides a map of how this survey is constructed. The top band summarizes construction of the 247-paper corpus through retrieval, LLM-assisted expansion, snowballing, screening, and structured annotation. The center introduces the lifecycle-based analytical lens used to trace risk from input, planning, decision, and tool execution to output, with memory, coordination, information flow, delegated authority, and persistence treated as cross-cutting concerns. This lens allows the review to examine how risks propagate through agent systems.
The lower half connects this lens to RQ1--RQ4 on modeling, threats, defenses, and evaluation, followed by the recurring synthesis themes of trust boundaries, privilege control, state management, and deployment assurance. Corpus construction and thematic synthesis thereby share the same analytical boundaries.

\subsection{Research Questions}\label{sec:rq}

This survey is guided by four research questions on LLM agent security.
\begin{itemize}[leftmargin=6mm]
\item \textbf{RQ1 (LLM Agent Security Scope and Modeling):} How can LLM agent security be scoped and modeled as an agentic systems security problem?
\item \textbf{RQ2 (Threat Surfaces and Attack Families):} What threat surfaces and attack families dominate current research on LLM agent security?
\item \textbf{RQ3 (Defense Strategies and Tradeoffs):} What defense mechanisms have been proposed to secure LLM agents, and what tradeoffs do they introduce?
\item \textbf{RQ4 (Evaluation Practices and Benchmark Gaps):} How is LLM agent security evaluated, and what gaps remain in existing benchmarks and empirical methodologies?
\end{itemize}

\begin{figure}[!htbp]
\centering
\includegraphics[width=0.85\textwidth]{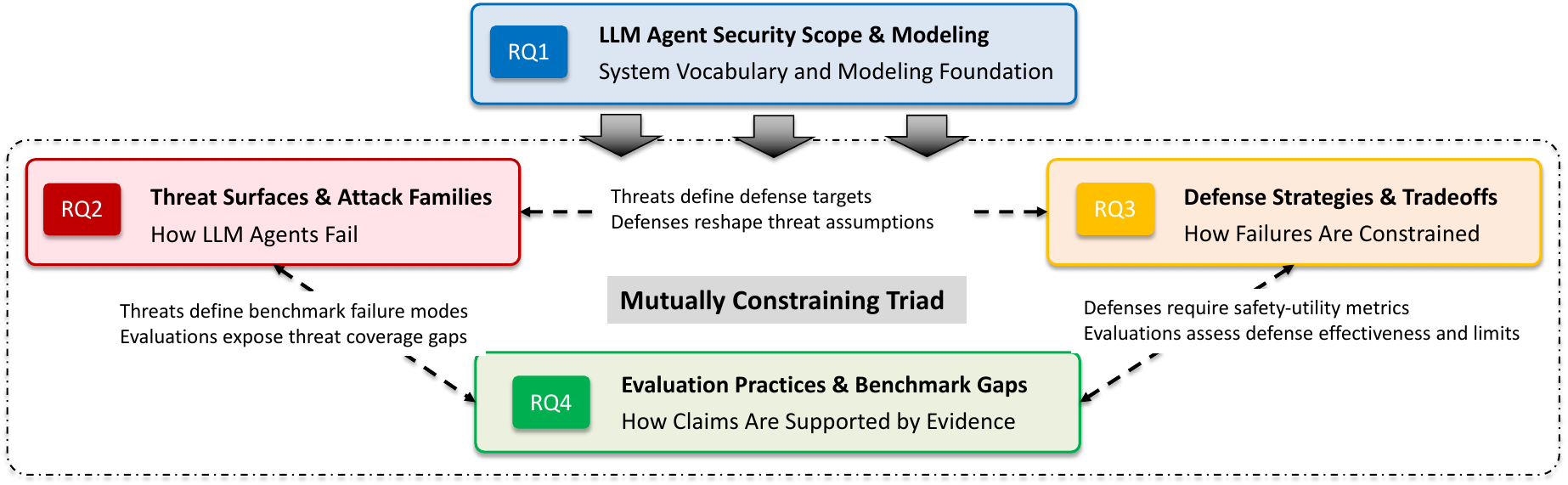}
\caption{Relationships of Research Questions}
\label{fig:rqs}
\end{figure}

As shown in \figref{fig:rqs}, RQ1 establishes the modeling foundation by defining what counts as LLM agent security and by locating security-relevant behavior within the agentic loop. This foundation is necessary because claims about attacks, defenses, and benchmarks all depend on a shared understanding of where information enters, where authority is exercised, and where state persists.
RQ2, RQ3, and RQ4 are more tightly intertwined. RQ2 identifies threat surfaces and attack families, but the meaning of those threats depends partly on what defenses assume and where evaluations can observe failure. RQ3 examines defense mechanisms, but a defense can only be interpreted against the threat paths it is meant to constrain and the metrics used to judge its safety-utility tradeoffs. RQ4 studies benchmarks and empirical practice, but benchmark coverage is meaningful only in relation to the threats and defenses that the benchmark exercises. In this sense, threats, defenses, and evaluations form a mutually constraining triad.
RQ1 provides the common systems vocabulary, while RQ2--RQ4 examine three interdependent views of the same security problem.

To answer RQ1, we annotate the lifecycle stages addressed by each paper and synthesize how untrusted information, delegated authority, and persistent state shape agent security. To answer RQ2, we compare threat surfaces, attack methods, and deployment scenarios to distinguish dominant attack families from emerging persistence and propagation risks.
For RQ3, we classify defenses by intervention point, protected asset, trust assumption, and reported tradeoffs, which allows us to assess both their prevalence and composability. For RQ4, we examine scenario coverage, benchmark reuse, metrics, and methodological realism, including whether evaluations capture long-horizon behavior, memory, coordination, privilege-sensitive action, utility, latency, and cost.

\subsection{Search Strategy}\label{sec:m1}

We construct the corpus through database retrieval, LLM-assisted candidate expansion, and citation snowballing \cite{kitchenham2007guidelines, wohlin2014snowballing}. The search window runs from January 1, 2023 to April 27, 2026, covering the field's emergence in late 2023 and rapid expansion after 2024.

The first stage searches ACM Digital Library, IEEE Xplore, Scopus, Web of Science, arXiv, and Google Scholar to cover archival venues, fast-moving preprints, and cross-disciplinary indexing. This combination is important because relevant studies appear across security, systems, software engineering, machine learning, and NLP, and are often easier to find through benchmark or framework names than stable subject headings. We combine LLM, agent, and security terms while adapting field selectors and phrase syntax to each interface.
The second stage uses GPT-5.4 with web search as a bounded retrieval assistant to suggest lexically distant papers, query variants, benchmark aliases, and missing topic branches. It does not determine inclusion, and every candidate is manually verified. The third stage applies backward and forward snowballing from representative seeds, benchmarks, and surveys \cite{wohlin2014snowballing, deng2025ai, he2026emerged, yu2025survey, zhu2025survey}. The complete query, adaptation rules, expansion procedure, and prompt appear in \appref{sec:app-search}.

\subsection{Literature Selection}\label{sec:m2}

We apply a staged eligibility protocol to construct the final corpus \cite{kitchenham2007guidelines}. It begins with 209 database records, 42 from LLM-assisted expansion, and 24 from snowballing. Title, abstract, and when necessary full-text screening excludes 24 of the 275 records. Normalization merges four duplicate versions or preprint-publication pairs, leaving 247 papers. Eligible studies substantively address LLM agent security and an explicit agentic surface such as planning, tools, external action, persistent state, runtime infrastructure, embodied interaction, MCP or skill interfaces, or multi-agent communication. We exclude generic prompt security without an agentic loop, indirectly related safety or governance work, autonomous-system studies without a central security focus, and capability-only benchmarks. \appref{sec:app-selection} gives the complete breakdown and boundary rationale.

\subsection{Literature Annotation and Analysis}\label{sec:m3}

We manually classify each paper for review synthesis rather than source-code analysis. The scheme records bibliographic metadata together with paper type, system setting, topic, scenario, threat model, attack surface, lifecycle stage, benchmarks, metrics, and attack and defense methods. These dimensions connect corpus-level distributions to the four research questions.

Paper type and system setting are single-label, while technical dimensions are multi-label. Three authors independently annotated all 247 papers, encoding multi-label categories as binary decisions. Agreement across the main fields reached Fleiss' $\kappa=0.86$ before harmonization. \appref{sec:app-annotation} reports operational boundaries, field-level values, and resolution procedures.

\section{Trend Observation}\label{sec:trend}

\begin{figure}[!htbp]
\centering
\includegraphics[width=0.9\linewidth]{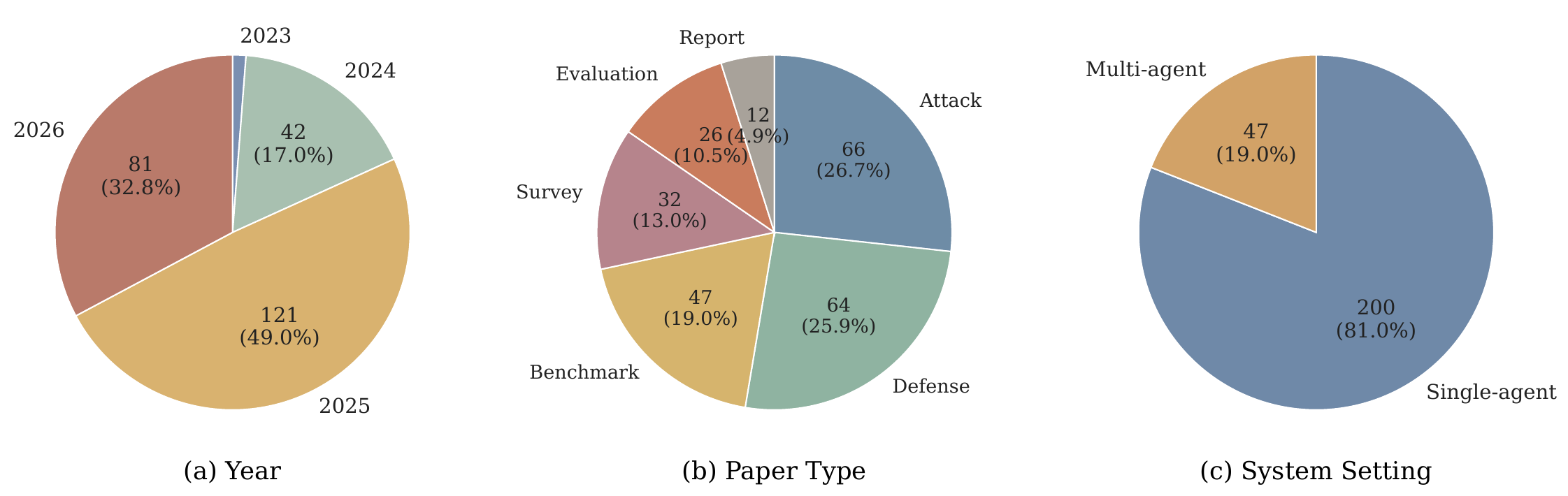}
\caption{Corpus Distributions by Year, Primary Paper Type, and System Setting}
\label{fig:trend}
\end{figure}

\figref{fig:trend} summarizes publication year, paper type, and system setting. Together, these distributions show rapid growth alongside unsettled venue structure, diverse research functions, and uneven architectural coverage.
The corpus grows from 3 papers in 2023 and 42 in 2024 to 121 in 2025, marking a shift from exploratory work to a concentrated research agenda. The 81 papers collected by April 27, 2026 already represent 32.79\% of the corpus. Although this partial-year count is not a final annual total, it indicates sustained activity after the 2025 surge. The compressed timeline also means that terminology, threat models, evaluation protocols, and deployment assumptions remain in flux, requiring cautious interpretation and version normalization.

ArXiv preprints account for 169 papers, or 68.42\% of the corpus, while the remainder spans machine learning, NLP, security, software engineering, robotics, AI ethics, and industry sources. This long tail shows that the field has not consolidated around a small group of archival venues and cautions against treating publication frequency as evidence of threat consequence, defense maturity, or deployment readiness. \appref{sec:app-trend} provides the normalized venue distribution and identity rules.
Attack and defense papers are the two largest categories with 66 and 64 papers. Together they comprise 52.63\% of the corpus, showing that mitigation has developed alongside threat discovery rather than only as a delayed response. The remainder includes 47 benchmark papers, 26 evaluation papers, 32 surveys, and 12 reports, indicating a visible measurement and synthesis layer. This diversity does not imply methodological maturity because studies still operationalize agent security through divergent threat models, environments, metrics, and attacker assumptions.
Single-agent systems account for 200 papers, or 80.97\%, while multi-agent systems account for 47, or 19.03\%. The evidence base therefore remains centered on individual tool-using, browsing, coding, and memory-augmented agents. Yet the multi-agent share rises from 9.52\% in 2024 to 23.97\% in 2025 and remains 17.28\% in the partial 2026 corpus. Multi-agent security is becoming a stable subarea concerned with communication, delegation, role separation, coordination failure, and cross-agent propagation rather than a peripheral niche.
Taken together, the field is temporally compressed, preprint-heavy, functionally diverse, and architecturally uneven. Attacks, defenses, benchmarks, evaluations, surveys, and reports all occupy visible portions of the corpus, while single-agent systems still dominate its architectural coverage. These properties motivate analysis beyond isolated attack and defense categories.

These distributions also require caution. Preprints may change before publication, primary paper-type labels simplify studies that serve several functions, and single-agent dominance may underrepresent delegation and coordination risks. These limitations reinforce organization around the agentic loop rather than venues, paper types, or application domains alone.

\section{RQ1. LLM Agent Security Scope and Modeling}\label{sec:rq1}
RQ1 asks how LLM agent security should be scoped and modeled as an agentic systems security problem. We use the \emph{agentic loop} that transforms language inputs into plans, decisions, tool invocations, state updates, and external effects. This scope is broader than prompt security but narrower than generic AI safety. It does not treat every failure of an LLM-enabled system as an agent-security problem. Instead, it focuses on the software structure through which information becomes control, control becomes action, and action produces durable consequences.

We represent an LLM agent with the semi-formal tuple $A=\langle I, P, D, T, M, O, C\rangle$, where $I$ denotes inbound context and observations, $P$ planning over candidate trajectories, $D$ commitment to a concrete next action or delegation step, $T$ tool or environment execution, $M$ transient or persistent state, $O$ externally visible outputs and side effects, and $C$ coordination channels with monitors, humans, or peer agents. Security-relevant behavior emerges not from any one element in isolation, but from the flows among them. Low-authority content in $I$ can distort $P$, change $D$, trigger privileged $T$, poison $M$, or propagate through $C$. This representation is intentionally lightweight, but it makes explicit where trust, authority, and persistence interact.

The empirical center of the corpus is consistent with this framing. The most frequent research topics are \textit{Tool-use Security} with 156 papers, \textit{Runtime Defense} with 88, \textit{Prompt Injection Security} with 75, \textit{Multi-agent Security} with 63, and \textit{Memory Safety} with 32. These topic frequencies are revealing. They show that the literature is already oriented around control, capability use, runtime mediation, coordination, and state integrity, even when individual papers use different labels or enter the problem from different application domains. Survey and taxonomy papers make the same move from prompt-level risk to agent-level control concerns \cite{liu2024personal, deng2025ai, he2025security, he2026emerged, ji2024navigating, yu2025survey, liu2025comprehensive, han2025survey, zhu2025survey, raza2026trism, minnich2025systematization}. Related governance and systems analyses further emphasize delegated authority, state, assurance, and deployment context \cite{wang2025model, liu2025towards, lam2025forewarned, yan2025protecting, gao2025four, ma2025safety, ferrag2026prompt, guan2025sok, namiot2026prompt, yan2026agent, ji2026agentic, yu2026sok, zhang2026what, song2026framework, liu2026thinker, tang2026security, bhosale2026dark, luo2026towards, shin2025seven, jha2026human, arremsetty2026security}. Deployment reports, system cards, and operational advisories point in the same direction from the practice side \cite{flynn2025lessons, openai2025operator, openai2025chatgpt, anthropic2024developing, deepmind2025gemini, deepmind2026gemini, team2026running, team2026openclaw, vavra2026how, openai2026gpt, anthropic2025claude, github2026openclaw}.

The lifecycle distribution supports this model. Among papers with lifecycle annotations, \textit{Planning} appears in 227 papers, \textit{Input} in 225, \textit{Tool Execution} in 209, \textit{Decision} in 166, and \textit{Output} in 151. Memory and coordination are less frequent but still substantial, with \textit{Memory} appearing in 82 papers, \textit{Monitoring} in 58, \textit{Inter-agent Comms} in 35, and \textit{Coordination} in 20. The field thus studies repeated, stateful control with external consequences rather than isolated prompt failures.
On this basis, the lifecycle view is useful not only descriptively, but also analytically. It provides a common unit of comparison across attacks, defenses, and benchmarks. It also explains why seemingly different problems, such as prompt injection, memory poisoning, unsafe delegation, and privilege abuse, can be understood as different disruptions of the same loop. \tabref{tab:rq1-lifecycle} identifies the recurrent dimensions that anchor this conceptual model.

\begin{table}[!htbp]
\caption{Core Dimensions for Modeling LLM Agent Security}
\scriptsize
\label{tab:rq1-lifecycle}
\centering
\setlength{\tabcolsep}{4pt}
\renewcommand{\arraystretch}{1.08}
\begin{tabularx}{\textwidth}{L{0.17\textwidth}C{0.05\textwidth}Y}
\toprule
\textbf{Dimension} & \textbf{Papers} & \textbf{Why It Matters} \\
\midrule
\textbf{Lifecycle Stage} &  & \\
Planning & 227 & Distorted decomposition and step sequencing are central to how harm materializes. \\
Input & 225 & Untrusted content enters through multiple channels and therefore shapes the control problem itself. \\
Tool Execution & 209 & Many consequential failures are realized only when agents cross capability boundaries. \\
Decision & 166 & Local action commitments connect reasoning to observable behavior. \\
Memory & 82 & Persistent state turns one-shot compromise into delayed or recurring compromise. \\
\midrule
\textbf{Threat Surface} &  & \\
User Prompts & 82 & Direct instructions still matter, but they no longer define the whole risk picture. \\
Web Content & 55 & Browsing environments continuously mix task-relevant data with hidden control signals. \\
Tool Outputs & 54 & Tool responses are often overtrusted and thus become a mediated attack path. \\
Retrieved Content & 37 & Retrieval can serve as both grounding evidence and attack carrier. \\
Memory / Scratchpads & 25 & Internal state is both an asset and an attack surface once it is reused. \\
\bottomrule
\end{tabularx}
\end{table}

\subsection{Lifecycle Stages and Major Threat Surfaces}\label{sec:rq1-2}

The lifecycle view becomes more informative when it is paired with threat surfaces. Lifecycle stages explain \emph{when} a risk acts. Threat surfaces explain \emph{where} it enters or what asset it touches. The surface distribution shows that direct prompting is only one part of the security story. \textit{User Prompts} are the single most frequent threat surface with 82 papers, but \textit{Web Content} appears in 55 papers, \textit{Tool Outputs} in 54, \textit{Retrieved Content} in 37, and \textit{Files / Code}, \textit{Planning Loop}, \textit{Memory / Scratchpads}, and \textit{Inter-agent Channels} each in at least 25. This pattern indicates that the field is already centered on mediated and internalized control paths rather than on user prompts alone.

These surface labels are deliberately narrower than ``untrusted text''. \textit{Web Content} denotes content fetched from open browsing environments. \textit{Retrieved Content} denotes evidence returned by retrieval modules or search indices. \textit{Tool Outputs} denotes results produced by an invoked tool or API. \textit{Files / Code} denotes local artifacts that may be read, executed, or modified. \textit{Memory / Scratchpads} denotes state generated or retained by the agent itself. Distinguishing these surfaces matters because they differ in authority assumptions, persistence, and feasible enforcement points.

Taken together, these surface counts also clarify what is distinctive about agent security. In many current architectures, the same natural-language substrate contains user goals, retrieved evidence, tool feedback, intermediate plans, summaries, and delegated messages. The central modeling challenge is therefore not only malicious-input detection. It is the preservation of boundaries between data and control, between low-authority observations and high-authority instructions, and between transient context and durable state. Once those boundaries are blurred, contamination may move from one stage to another without any explicit privilege transition.

The distinction among \textit{Planning}, \textit{Decision}, and \textit{Tool Execution} is likewise operational. \textit{Planning} captures decomposition, intermediate reasoning, or trajectory generation. \textit{Decision} captures commitment to one next step or delegation target. \textit{Tool Execution} captures the privileged act itself. In ReAct-like systems these steps may be tightly interleaved, but separating them clarifies whether a paper studies reasoning corruption, action selection, or authority crossing.

Three modeling principles follow from this observation. The first is \textit{data-control ambiguity}, since agents routinely consume text that is semantically relevant to the task but normatively untrusted. The second is \textit{delegated authority}, where attackers gain leverage because the agent can browse, execute, store, or communicate under permissions that the attacker does not directly own. The third is \textit{persistence and propagation}, where harm is often delayed because dangerous content is reused later through memory or coordination channels. The value of the lifecycle-by-surface model is precisely that it makes these cross-stage interactions explicit rather than treating them as disconnected edge cases.
The full lifecycle-by-surface matrix in \appref{sec:app-rq1-matrix} reports the underlying co-occurrence counts and distinguishes direct, mediated, persistent, and coordination pathways.

\subsection{Implications for Modeling LLM Agent Security}\label{sec:rq1-3}

The practical implication of this model is that agent security should be analyzed in terms of \emph{transitions} rather than components alone. Inputs, plans, tools, memory, and coordination channels are all important, but the most consequential failures usually occur when one kind of state is allowed to transform into another without adequate mediation. Untrusted content becomes especially dangerous when it is reinterpreted as a planning constraint, when a tentative plan becomes an executable decision, or when a stored trace is later reused as trusted context. This transition-centered view helps explain why agent security often looks different from both conventional application security and prompt-only model safety. The key question is not only what information the system has seen, but what the system is allowed to \emph{do next} because it has seen it.

This perspective also sharpens the architectural meaning of trust boundaries. In a conventional software pipeline, developers often know where data validation ends and privileged execution begins. In agentic systems, those boundaries are easier to blur because natural-language context may simultaneously contain user goals, retrieved evidence, tool feedback, intermediate reasoning, and delegated instructions. The tuple therefore provides a way to ask more precise design questions about which channels into $I$ should be treated as low-authority, which transitions into $P$ and $D$ require source-aware interpretation, which actions at $T$ deserve explicit capability checks, which state written into $M$ can be reused later, and which messages moving through $C$ may propagate authority rather than merely information. Once stated in this form, several seemingly separate issues in the literature become variations of the same modeling problem.

This modeling choice helps explain why the dominant threats manipulate control flow rather than merely content, why many promising defenses intervene at capability boundaries or runtime mediation points, and why evaluation becomes less satisfactory once persistence, coordination, and externally consequential action move to the foreground. It also provides a common language for comparing different deployments. A browsing attack, a coding-agent compromise, a memory-poisoning case, and a multi-agent contagion event can all be compared by where they enter the loop, which transitions they exploit, and where consequences are realized. The tuple thus locates attack entry, defensive intervention, and benchmark coverage within one model.

\subsection{Answer to RQ1}\label{sec:rq1-answer}

LLM agent security should be modeled through interactions among information flow, delegated authority, and persistent state across the agentic loop. More specific than calling the problem ``system-level'', this formulation unifies strands of the literature that might otherwise appear disconnected. Prompt injection is often a planning and execution problem, memory poisoning is a delayed control-flow problem, and multi-agent coordination is a trust and propagation problem.

\section{RQ2. Threat Surfaces and Attack Families}\label{sec:rq2}

RQ2 asks which threat surfaces and attack families dominate current research on LLM agent security. The corpus points in two directions at once. At the level of volume and visibility, the field is still centered on prompt injection and its tool-mediated variants. At the level of conceptual development, however, the threat model is expanding towards persistence, authority abuse, and propagation. Put differently, the literature is moving from single-step prompt compromise towards longer-horizon and networked forms of failure, even though prompt injection remains the dominant organizing problem.
The papers behind this characterization span mediated prompt injection in web and tool settings \cite{kang2024llm, xiao2024wipi, fu2024imprompter, liu2025autohijacker, wang2025webinject, lee2025chatinject, xie2025queryipi, xiao2025agenttypo, wang2026webweaver}, coding and computer-use compromise \cite{kang2024llmb, nakash2025breaking, liu2025red, xie2025red, lo2025your, patlan2025real, pei2026your, feng2026your}, and memory, storage, or tool-chain corruption \cite{li2024agentpoison, tiwari2024prompt, dong2025memory, he2025memorygraft, li2026mcp, herrador2026spaiware, lim2026adaptools, gong2026maltool, torr2026skillject, song2026storage, cheng2026skillattack, huang2026skilltrojan}. Multi-agent and propagation-centric attacks form a parallel branch \cite{kang2026teams, shmatikov2025multi, wang2025manipulating, shao2025collaborative, wang2026shadows, bibi2026omni, dong2026zombie}. Other studies extend the threat model to embodied backdoors, delayed triggers, social-engineering variants, and policy-evasion attacks \cite{zhu2025can, wang2024badagent, zhang2024badrobot, zhang2025breaking, xu2024poex, li2025udora, sun2026prompt, chen2025beyond, zhang2025eva, he2025unveiling, chen2025agents, yair2025invitation, ji2025automatic, xue2025mind, song2025agentvigil, lam2026beyond, kantarcioglu2026bypassing, choi2026overthinking, westrum2026silent, arp2026post, geng2026white, men2025troublemaker, guo2025corba, cheung2025ip, he2025demonstrations, yan2026attack, liu2025tipping, cao2026cia, talukder2026semantic}. Across these works, the most useful distinction is not simply between different attack names, but between different \emph{propagation logics}, including mediated entry through untrusted content, delayed reactivation through stored state, and amplification through inter-agent communication.

\subsection{Prompt Injection and Control-Flow Hijacking}\label{sec:rq2-1}

\tabref{tab:rq2-threats} summarizes the threat families that appear most prominently in the corpus. In particular, prompt injection, malicious tool mediation, and data exfiltration often share the same upstream control problem even when their immediate consequences differ across real deployment settings.

Prompt injection is the most visible and most structurally influential threat family in the corpus. In the threat-model field, \textit{Prompt Injection} appears in 142 papers and \textit{Indirect Prompt Injection} in 86. In the attack-method field, \textit{Indirect Prompt Injection} is again the largest normalized attack method, followed by \textit{Prompt Injection}, with both clearly ahead of later categories such as \textit{Memory Poisoning}, \textit{Backdoor Triggering}, and \textit{Data Exfiltration}. The meaning of this concentration is not merely that one topic is fashionable. It indicates that the field has largely identified control-flow hijacking as the defining weakness of current agents.
What makes this threat family especially important in agentic settings is that it is often mediated rather than direct. Direct user-side prompt injection remains relevant, but the attack-surface statistics show that the more characteristic risk comes from indirect sources, such as \textit{Web Content}, \textit{Tool Outputs}, and \textit{Retrieved Content}. These channels allow malicious instructions to be embedded in content that is task-relevant but not authority-bearing. Once such content is ingested into the reasoning context, the agent may treat it as actionable rather than merely informative. The security problem is therefore not simply prompt abuse. It is the failure to preserve instruction hierarchy and source legitimacy inside the agentic loop.

The concentration on prompt injection is visible across the most active deployment scenarios. In \textit{Web Browsing}, \textit{Prompt Injection} appears 71 times and \textit{Indirect Prompt Injection} 44 times. In \textit{Software Engineering}, the same two categories remain dominant with 32 and 16 occurrences. Even in \textit{Finance Tools}, \textit{Prompt Injection} and \textit{Indirect Prompt Injection} remain the two leading threats. The recurring lesson is that prompt injection is not just one attack class among many in the current literature. It is the main mechanism through which the corpus understands how untrusted content becomes unsafe control. This should still be read as a statement about \emph{research concentration} rather than a definitive ranking of all deployment risks.
In tuple terms, prompt injection and indirect prompt injection primarily target $I \rightarrow P$ and $I \rightarrow D$. Malicious tools and data exfiltration become operational at $D \rightarrow T$ and $T \rightarrow O$, memory poisoning targets $I/T \rightarrow M$ and later $M \rightarrow P$, and coordination failures propagate through $C \rightarrow P$ and $C \rightarrow D$. This mapping is not meant as a complete formalization, but it helps align attack families with the same system model used later for defenses and benchmarks.

\begin{table}[!htbp]
\caption{Dominant Threat Families and Propagation Logic}
\scriptsize
\label{tab:rq2-threats}
\centering
\setlength{\tabcolsep}{4pt}
\renewcommand{\arraystretch}{1.04}
\begin{tabularx}{\textwidth}{L{0.17\textwidth}C{0.05\textwidth}L{0.40\textwidth}Y}
\toprule
\textbf{Threat Family} & \textbf{Count} & \textbf{Common Surfaces and Scenarios} & \textbf{Main Propagation Pattern} \\
\midrule
Prompt Injection & 142 & User prompts and conversation context, especially in web browsing, collaborative settings & Local instruction manipulation that redirects planning or action selection \\
\addlinespace[3pt]
Indirect Prompt Injection & 86 & Web content, tool outputs, retrieved content, and files in browsing, finance, and coding agents & Hidden instruction injection through mediated content channels \\
\addlinespace[3pt]
Unsafe User Instructions & 43 & Underspecified or harmful task requests in browsing, coding, and embodied settings & Harmful task acceptance or compliance under weak task framing \\
\addlinespace[3pt]
Malicious Tools & 34 & Tool outputs, metadata, or plugin/API interfaces in coding, healthcare, and finance agents & Misleading tool-mediated reasoning or unsafe external invocation \\
\addlinespace[3pt]
Data Exfiltration & 31 & Tools, files, retrieved content, and external services in email, finance, and healthcare workflows & Leakage of sensitive information through tool use or state exposure \\
\addlinespace[3pt]
Memory Poisoning & 24 & Memory stores, summaries, scratchpads, and retrieved state in long-horizon or collaborative agents & Delayed corruption that reappears in later plans and actions \\
\addlinespace[3pt]
Coordination Failures & 14 & Inter-agent channels, roles, and delegation logic in multi-agent collaboration & Amplification of local failure through delegation and message passing \\
\bottomrule
\end{tabularx}
\end{table}

\subsection{State Integrity and Multi-Agent Propagation}\label{sec:rq2-2}

Prompt injection remains dominant, but it is no longer sufficient to characterize the full threat landscape. The corpus increasingly points towards threats that exploit persistence and propagation. These threats are especially visible in work on memory integrity and multi-agent coordination.

State integrity is the clearest example. \textit{Memory Poisoning} appears in 24 papers as a threat model, while \textit{Memory Safety} appears in 32 papers as a research topic. These numbers are still much smaller than those of prompt injection, but their conceptual weight is substantial. Once an agent stores notes, summaries, or long-lived context, contamination may survive the original interaction and reappear during a later planning episode. The security question is no longer only whether a malicious instruction was seen, but whether the system can remember, trust, forget, quarantine, and revoke correctly over time. Memory should therefore be understood not as an auxiliary module, but as a primary security surface in long-horizon agents.

Multi-agent systems expose a parallel expansion from local failure to distributed failure. Although only 47 corpus papers are explicitly multi-agent, the annual distribution shows a clear trend. The share of multi-agent papers rises from 9.52\% in 2024 to 23.97\% in 2025 and remains at 17.28\% in the partial 2026 corpus. This rise is consistent with the research-topic distribution, where \textit{Multi-agent Security} already appears in 63 papers. The field is therefore not merely adding more collaborative architectures. It is beginning to recognize coordination as a source of distinct security semantics.

The threat pattern of collaborative settings confirms this interpretation. In \textit{Multi-agent Collaboration}, \textit{Prompt Injection} still appears 44 times, but it co-occurs with \textit{Memory Poisoning} 14 times and \textit{Coordination Failures} 14 times. This means that multi-agent security cannot be reduced to repeating single-agent risks at larger scale. It also includes message integrity, role confusion, failure amplification, and topology-sensitive contagion.

\begin{figure}[!htbp]
\centering
\subfigure[Mediated Injection Through External Content]{\includegraphics[width=0.32\textwidth]{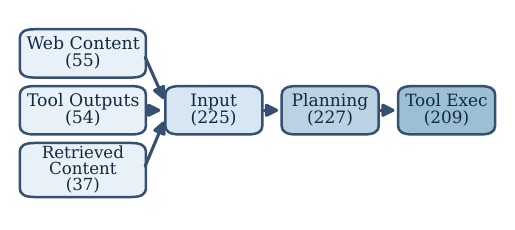}}
\hfill
\subfigure[Delayed Activation Through Poisoned Memory]{\includegraphics[width=0.32\textwidth]{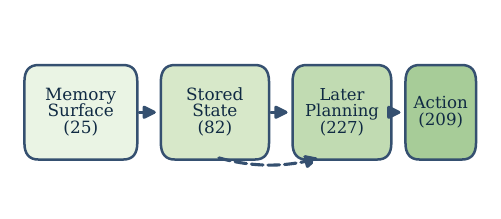}}
\hfill
\subfigure[Message-Level Spread Across a Multi-Agent Topology]{\includegraphics[width=0.32\textwidth]{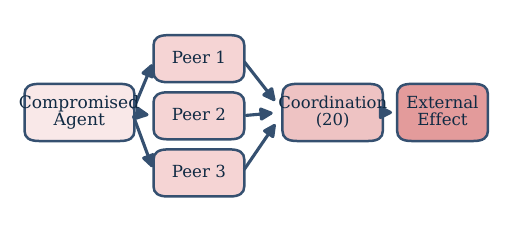}}
\vspace{-2mm}
\caption{Representative Propagation Patterns of LLM Agent Attacks}
\label{fig:rq2-propagation}
\end{figure}

\figref{fig:rq2-propagation} makes this shift concrete through three propagation patterns that recur across the corpus. Panel (a) shows mediated entry through external content, where \textit{Web Content} with 55 papers, \textit{Tool Outputs} with 54, and \textit{Retrieved Content} with 37 converge on the dominant operational chain of \textit{Input} (225), \textit{Planning} (227), and \textit{Tool Execution} (209). Panel (b) shows delayed state activation, with \textit{Memory / Scratchpads} appearing in 25 papers while the \textit{Memory} stage appears in 82 and \textit{Planning} in 227, illustrating how contaminated state can persist and later re-enter the planning process before action is taken. Panel (c) shows inter-agent spread, where compromise can be relayed through peers into \textit{Coordination} (20) and then external effect, making message-passing structure a security concern rather than an incidental implementation detail.

Threat frequency alone is not enough to characterize the field. Consequence asymmetry matters just as much. \textit{Web Browsing} with 93 papers and \textit{Software Engineering} with 63 define the empirical center of gravity of current work, which helps explain why prompt injection and tool-mediated abuse dominate the literature. However, the highest-consequence deployments are not always the most frequently studied. \textit{Finance Tools} with 34 papers, \textit{Healthcare Assistance} with 28, and \textit{Embodied Robotics} with 28 expose agents to stronger confidentiality, integrity, and physical-world consequences. In healthcare settings, for example, \textit{Memory Poisoning}, \textit{Data Exfiltration}, and \textit{Malicious Tools} all appear as visible risks. In embodied settings, \textit{Jailbreak Prompts} and unsafe instructions remain unusually salient. The implication is that survey conclusions should distinguish what is most common from what is potentially most costly.

The prominence of software engineering scenarios is particularly informative because coding agents combine several of the surfaces identified above within one development workflow. Repository content, issue descriptions, generated code, package metadata, and tool responses can all enter the agent as context, while file modification, shell execution, dependency installation, and secret access give compromised decisions concrete effects \cite{kang2024llmb, nakash2025breaking, xie2025red, lo2025your}. In these settings, prompt injection is not only a model-input problem. It can become a tool-chain and development-environment problem when untrusted project artifacts influence actions performed under the agent's delegated permissions. The reviewed evidence therefore supports threat models that follow provenance and authority across the full path from repository content to executable action.

\subsection{Answer to RQ2}\label{sec:rq2-answer}

Prompt injection and indirect control-flow hijacking dominate current research because untrusted prompts, web content, retrieved content, and tool outputs can influence the planning and execution chain. Threat models are nevertheless broadening towards persistent state corruption, authority abuse, and inter-agent propagation. This shift moves the unit of analysis from isolated malicious inputs to cross-stage compromise, delayed effects, and consequences shaped by deployment context.

\section{RQ3. Defense Strategies and Tradeoffs}\label{sec:rq3}

RQ3 asks what defense mechanisms have been proposed to secure LLM agents and what tradeoffs they introduce. The corpus suggests that defense is no longer a secondary reaction to attack discovery. It is now a substantial and rapidly growing part of the field. At the same time, the defense space remains fragmented, with few techniques reaching the status of a stable and reusable stack. The central challenge is therefore no longer whether defenses exist, but whether the current defense landscape can be organized into a coherent engineering logic that matches the propagation pathways identified in RQ2.

The need for a compositional account is visible in the range of mechanisms now being studied. The defense literature covers source-ordering and prompt-handling mechanisms \cite{yang2024plug, bagdasarian2024airgapagent, squicciarini2025task, tramer2025defeating, lee2025prompt, li2025llm, shmatikov2025breaking, beutel2024instruction, kiciman2024defending}, runtime monitoring and guard-agent architectures \cite{li2024guardagent, wang2025g, wang2025melon, gibbons2025rtbas, sun2025agentspec, nitarotaru2026ace, laboratory2026agentdog, zhao2026camels, sun2026clawguard}, privilege control, safer tool mediation, and containment \cite{iqbal2025isolategpt, popa2025miniscope, anagnostopoulos2026mcp, li2026taming, shao2026toolsafe, hu2026agentsentry, rao2026authenticated, li2026openclaw}, as well as memory, provenance, protocol, or communication protection \cite{shokri2025firewalls, liu2025os, wang2025advancing, song2025progent, narayan2025securing, saxe2025llamafirewall, jing2025mcip, zanellabeguelin2025securing, volhejn2025design, xiao2025drift, dhar2025securing, wu2025agentarmor, zhan2025blocka2a, an2025ipiguard, wang2025memguard, wagner2025better, salvaneschi2025securing, moaty2025toward, vainshtein2026agentrim, liu2026defense, mishra2026memory, zhang2026agentsys, lim2026icon, castromaldonado2026semantic, qin2026attriguard, li2026tamingb, hu2026trinityguard, wen2025agentsafe, zhu2025who, zhou2026spark, wang2025blindguard, liu2025advevo, sun2025monitoring, ji2026beyond, rahimi2026agenticcyops, amiri2026information, deng2026planguard, zhang2026auditing}. The diversity of this literature is encouraging, but it also creates a familiar problem. The field has many defensive components without yet having a strong account of how those components should compose under realistic and heterogeneous threat assumptions.

\subsection{Defense Families Along the Agentic Loop}\label{sec:rq3-1}

The strongest way to organize the defense space is by intervention point along the agentic loop. This framing clarifies both what current defenses protect and where their limits lie. It also helps separate defenses that primarily influence interpretation from those that constrain action or contain damage after interpretation has already failed.

The first family concerns source handling and instruction ordering. Methods such as \textit{Instruction Hierarchy} and \textit{Guardrails} try to reduce the chance that low-authority or malicious content is treated as a legitimate control signal. Their appeal lies in their relatively low deployment cost and architectural simplicity. However, their protection remains fundamentally model-mediated. They still depend on the model correctly interpreting source precedence in noisy, multi-source contexts, which means they improve robustness without fully relocating trust away from the model.

The second family concerns runtime scrutiny. \textit{Runtime Monitoring}, \textit{Policy Enforcement}, and \textit{Anomaly Detection} all intervene after content has already entered the loop but before harmful consequences are fully realized. These methods are attractive because they match the temporal structure of agent behavior more closely than prompt-only defenses do. They can inspect plans, tool calls, or execution traces near the point of consequence realization. Their cost is overhead, since they add latency, increase system complexity, and often rely on secondary reasoning or classification modules whose own robustness may be uncertain.

The third family concerns capability control and containment. \textit{Access Control}, \textit{Information-flow Control}, \textit{Context Isolation}, and \textit{Sandboxing} aim to reduce what compromised reasoning is allowed to influence. This family is especially important because it targets authority boundaries rather than only input quality. In many respects, these methods are the closest analogues to mature software security mechanisms. Their limitation is not conceptual weakness but engineering difficulty, because fine-grained permissions, trusted mediation, provenance-aware separation, and practical isolation are harder to maintain in flexible natural-language systems than in conventional software stacks.

\tabref{tab:rq3-defenses} summarizes the main defense families that recur in the corpus together with the engineering assumptions they make. The table reveals an important structural feature of the field. Even the most recurrent techniques are not yet dominant in a strong sense, and the families protect different assets under different assumptions. The current literature has ingredients, but it has not yet converged on a widely accepted security stack or on a stable mapping from threat model to a workable set of necessary defensive layers.

\begin{table}[!htbp]
\caption{Defense Families, Tuple Focus, and Engineering Tradeoffs}
\tiny
\label{tab:rq3-defenses}
\centering
\setlength{\tabcolsep}{3.5pt}
\renewcommand{\arraystretch}{1.18}
\begin{tabularx}{\textwidth}{L{0.16\textwidth}L{0.12\textwidth}Y Y}
\toprule
\textbf{Defense Family} & \textbf{Tuple Focus} & \textbf{Primary Protection Goal} & \textbf{Main Tradeoff or Failure Mode} \\
\midrule
Input-trust management & $I \rightarrow P$, $I \rightarrow D$ & Prompt and indirect-prompt injection with control-flow integrity at mixed-trust inputs & Relies on the model or wrapper to preserve source distinctions under paraphrase, composition, or tool-mediated restatement \\
\addlinespace[3pt]
Runtime monitoring and guard agents & $P \rightarrow D$, $D \rightarrow T$, $T \rightarrow O$ & Unsafe plans, unsafe tool calls, and policy violations near the point of consequence realization & Adds latency and can over-block benign behavior, and depends on trajectory observability and policy quality \\
\addlinespace[3pt]
Access control and least privilege & $D \rightarrow T$, $T \rightarrow O$ & Privilege escalation, malicious tools, and unsafe capability use & Requires workable permission granularity and framework support for capability scoping \\
\addlinespace[3pt]
Information-flow and state isolation & $I \leftrightarrow M$, $M \rightarrow P$, $T \rightarrow M$ & Memory poisoning, cross-context leakage, and data/control confusion & Harder to sustain in long-horizon and collaborative settings without provenance or typed interfaces \\
\addlinespace[3pt]
Execution containment & $T \rightarrow O$ & Post-compromise damage, unsafe code execution, and external side effects & May impose engineering cost, compatibility limits, or reduced autonomy \\
\addlinespace[3pt]
Topology-aware multi-agent containment & $C \rightarrow P$, $C \rightarrow D$, $C \rightarrow C$ & Cross-agent contagion, coordination failures, and role confusion & Depends on visibility into messages or topology and remains weak against covert collusion channels \\
\bottomrule
\end{tabularx}
\end{table}

Representative work helps explain why a pure count-based synthesis is insufficient. Instruction hierarchy and related prompt-handling ideas are lightweight and easy to retrofit \cite{beutel2024instruction, tramer2025defeating}, but they continue to rely on the model's own ability to separate control from data. Guard-agent and runtime-policy approaches shift enforcement closer to the point of consequence realization \cite{li2024guardagent, sun2025agentspec, laboratory2026agentdog}, but they incur additional model calls, false-positive tradeoffs, and domain-specific policy design. Access-control and information-flow approaches offer a stronger systems-security logic because they constrain what compromised reasoning is allowed to affect \cite{popa2025miniscope, anagnostopoulos2026mcp, li2026taming, amiri2026information, dhar2025securing, zhang2026agentsys}, yet they demand explicit capability models, provenance signals, and engineering effort that many current agent frameworks do not yet expose by default. Finally, sandboxing and topology-aware multi-agent defenses are often the most robust against catastrophic failure \cite{iqbal2025isolategpt, li2026openclaw, zhan2025blocka2a, hu2026trinityguard, zhang2026auditing}, but they are also the most deployment-specific and infrastructure-heavy.
Through the tuple lens, current defenses cluster most strongly around $I \rightarrow P$ and $D \rightarrow T$. That is, they try either to prevent untrusted content from being interpreted as control or to restrict what chosen actions may do. By contrast, protections centered on $M$ and $C$ remain thinner and less standardized, which is one reason why memory-heavy and multi-agent deployments still lack stable defensive baselines.

\subsection{Enforcement, Isolation, and State Protection}\label{sec:rq3-2}

The lifecycle and threat analyses above reveal two higher-level defense patterns.
First, the most developed recurrent defenses in the corpus tend to protect authority boundaries rather than prompt boundaries alone. The strongest recurring ideas in the normalized corpus, especially access control, policy enforcement, and information-flow control, all try to constrain what contaminated reasoning is allowed to influence. This pattern suggests that the field is gradually translating classical security principles such as least privilege, mediation, and separation of duties into agentic settings rather than relying solely on better prompting discipline.
Second, defense coverage is still misaligned with the direction of the threat model. The field has already identified persistent memory corruption and coordination failures as important concerns, yet comparatively little defense work directly addresses provenance, revocation, inter-agent trust, or topology-aware containment. In other words, current defenses are densest around better-studied prompt-to-tool pathways and much thinner around the memory and coordination pathways which are still expanding.

\begin{figure}[!htbp]
\centering
\includegraphics[width=0.85\textwidth]{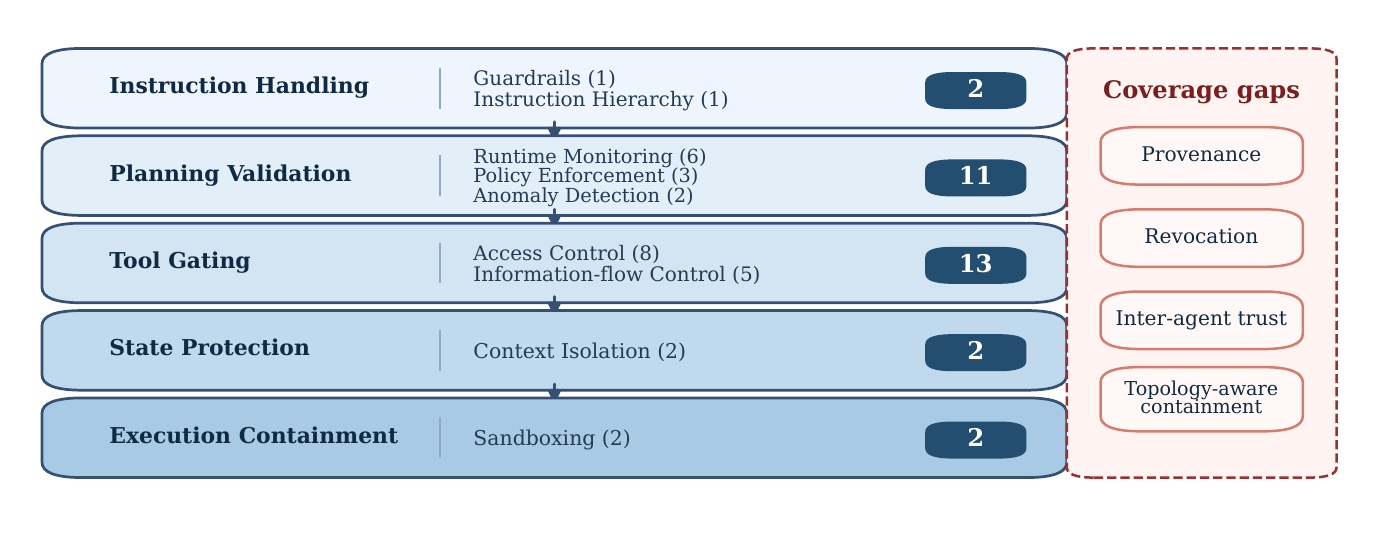}
\vspace{-5mm}
\caption{Layered Defense Stack and Coverage Gaps for LLM Agents}
\label{fig:rq3-stack}
\end{figure}

The resulting coverage asymmetry explains why a layered stack is preferable to any single dominant technique. Source ordering without capability control cannot prevent privilege abuse. Runtime monitoring without containment cannot fully limit external harm. Sandboxing without provenance or policy cannot address delayed state corruption. \figref{fig:rq3-stack} begins with \textit{Instruction Handling}, continues through \textit{Planning Validation} and \textit{Tool Gating}, and extends into \textit{State Protection} and \textit{Execution Containment}. The counts indicate that engineering effort concentrates on planning scrutiny and tool gating, while state protection and containment remain comparatively thin. Provenance, revocation, inter-agent trust, and topology-aware containment are consequently the least developed layers.

The corpus does not support a universal stack for every architecture, but it does support three scenario-level conclusions. Privilege control at $D \rightarrow T$ becomes broadly necessary once an agent can act externally. Memory protection becomes difficult to treat as optional once $M$ persists across tasks or sessions, and communication-aware containment becomes necessary when $C$ is an execution channel rather than a logging artifact. \appref{sec:app-rq3-scenarios} details this synthesis and its scope, trust, overhead, and utility tradeoffs across browser, coding, memory, and multi-agent systems.

\subsection{Answer to RQ3}\label{sec:rq3-answer}

Existing defenses span instruction handling, runtime monitoring, access control, information-flow control, and containment. They concentrate on transitions from input to planning and from decision to tool execution, while memory and coordination remain less protected. Because these mechanisms rely on different trust assumptions and impose different utility, latency, and engineering costs, no single technique provides a stable security stack. The field therefore needs a compositional account of which layers fit particular agent architectures, threats, and deployment stakes.

\section{RQ4. Evaluation Practices and Benchmark Gaps}\label{sec:rq4}

RQ4 asks how LLM agent security is evaluated and what gaps remain in existing benchmarks and empirical methodologies. The corpus reveals a familiar pattern. Evaluation activity is increasing quickly, but evaluation consensus is not. The field has produced a growing number of benchmarks and measurement-oriented studies, yet it still lacks a common evaluation backbone that can support strong cross-paper comparison, cumulative evidence, and realistic deployment claims.

The benchmark ecosystem is correspondingly broad. It includes prompt-injection and tool-use suites \cite{yuan2024r, zhan2024injecagent, tramer2024agentdojo, zhang2025agent, chen2024safeagentbench, reddy2025safearena, chaudhuri2025wasp, xiao2026agentdyn}, harmful-action and computer-use settings \cite{davies2025agentharm, andriushchenko2025os, wang2025ras, sun2025redteamcua, bagdasarian2025terrarium, ma2025shawshank, wang2026agentlab, ge2026clawsafety}, and web, email, office, finance, MCP, skill, and multi-agent evaluation suites \cite{liu2024exploring, liang2024cybench, wu2024riskawarebench, qiu2025evaluating, huang2024agent, lukacs2024hacksynth, zhang2025aeia, dvijotham2025doomarena, liang2025bountybench, wu2026crmarena, vainshtein2026maps, nakash2025effective, cherubin2025llmail, hooi2025vpi, sap2025openagentsafety, huang2025towards, radanovic2025benchmarking, chen2025mcpsecbench, wang2026mcptox, wang2025training, zeng2025safemind, pfister2025breaking, zhang2026mcp, liu2025securewebarena, gong2025wainjectbench, donaldson2026capture, liu2026assistant, andriushchenko2026skill, ji2026aciarena, ganu2025tamas, badumarfo2026agentleak}. Complementary evaluation and measurement studies further examine tool-use risk, deployment architecture, protocol-level exposure, shared-state contamination, and multi-agent trust tradeoffs \cite{hashimoto2024identifying, iqbal2024llm, ye2024toolsword, wang2024netsafe, cong2025agentguard, goldblum2025commercial, peigne2025multi, xiang2025how, luo2025master, gasmi2026bridging, hong2025mind, liang2025when, namiot2026breaking, rudolph2026too, shapira2026agents, zhan2026systems, fard2026are, kolter2026how, zhao2026no, wang2026your, witt2024secret, bansal2025sum, nambi2025exposing, wang2026attackeval, curvo2025traitors, das2026conscientia}. The remaining gap is a shared evaluation theory that connects threat surfaces, lifecycle stages, deployment scenarios, and reporting practice.

\subsection{Benchmark Landscape}\label{sec:rq4-1}

The benchmark landscape is now broad enough to be visible as a subfield in its own right. However, it is still too fragmented to function as a common standard. \textit{AgentDojo} is the most frequently reused benchmark with 30 occurrences, followed by \textit{InjecAgent} with 11. A second tier includes \textit{ASB} and \textit{SafeAgentBench} with 4 each, followed by \textit{R-Judge}, \textit{OSWorld}, \textit{MMLU}, \textit{CSQA}, \textit{MIMIC-III}, \textit{GSM8K}, and \textit{JailbreakBench}. Even the most reused benchmark accounts for only 12.82\% of all benchmark annotations in the corpus.

\begin{wraptable}{r}{0.35\textwidth}
\caption{Most Frequently Reused Benchmarks in the Reviewed Corpus}
\scriptsize
\label{tab:rq4-benchmarks}
\centering
\renewcommand{\arraystretch}{1.2}
\begin{tabular}{lrr}
\toprule
\textbf{Benchmark} & \textbf{Papers} & \makecell[c]{\textbf{Share}\space\textbf{(\%)}} \\
\midrule
AgentDojo & 30 & 12.82 \\
InjecAgent & 11 & 4.70 \\
MMLU & 5 & 2.14 \\
ASB & 4 & 1.71 \\
SafeAgentBench & 4 & 1.71 \\
R-Judge & 3 & 1.28 \\
OSWorld & 3 & 1.28 \\
JailbreakBench & 3 & 1.28 \\
\bottomrule
\end{tabular}
\end{wraptable}

The reuse distribution makes benchmark fragmentation difficult to miss. The field has several recognizable benchmarks, but no benchmark family yet spans the dominant threat surfaces, lifecycle stages, and deployment settings in a way that stabilizes empirical practice. Instead, current evaluation is organized around a patchwork of partially reusable suites, each answering a subset of questions. Such a patchwork can accelerate early experimentation, but it cannot yet support a stable comparative evidence base. \tabref{tab:rq4-benchmarks} indicates this weak concentration rather than benchmark influence. Because the benchmark field is multi-label, the percentage column uses total benchmark annotations as the denominator, and one paper may contribute to several benchmark counts.

Frequency alone does not show what each benchmark is actually good for. \textit{InjecAgent} and \textit{AgentDojo} are influential because they make prompt injection measurable in tool-integrated workflows, albeit under different assumptions about task dynamism and utility tradeoffs \cite{zhan2024injecagent, tramer2024agentdojo}. \textit{ASB} extends the space towards memory poisoning and backdoors \cite{zhang2025agent}. \textit{SafeArena} and \textit{WASP} emphasize unsafe task acceptance and web-agent browsing risk \cite{reddy2025safearena, chaudhuri2025wasp}, while \textit{OS-Harm} targets unsafe computer-use actions \cite{andriushchenko2025os}. Newer suites such as \textit{AgentDyn} and \textit{AgentLeak} push towards more open-ended or multi-agent settings \cite{xiao2026agentdyn, badumarfo2026agentleak}. In tuple terms, most reusable benchmarks still concentrate on $I$, $P$, $D$, and $T$, while $M$ and especially $C$ remain much less consistently exercised.

\subsection{Metrics and Sources of Fragmentation}\label{sec:rq4-2}

Metric practice reveals a second and equally important imbalance. \textit{Attack Success Rate} appears in 129 papers, far ahead of \textit{Success Rate} with 59 and \textit{Accuracy} with 41. By contrast, more deployment-sensitive metrics are much less common. \textit{Utility} appears in only 22 papers, \textit{Latency} in 8, and \textit{Cost} in 7. Even measures that are essential for operational tradeoffs, such as \textit{Refusal Rate} and \textit{False Positive Rate}, appear only 13 times each across the corpus.

The metric imbalance matters because LLM agent security is not a single-objective problem. A defense that reduces attack success by refusing every action is safe only in a trivial sense. Conversely, an agent that maintains task success while silently leaking data or corrupting memory is not secure. Much of the current literature therefore evaluates whether systems can be broken, but less often whether they remain governable, useful, and economically deployable under realistic controls. In that sense, many current benchmark practices still resemble vulnerability discovery more than deployment assurance under realistic use.

Fragmentation is compounded by differences in task construction and reporting practice. Different studies assume different tools, different attacker knowledge, different environment realism, and different utility baselines. Some focus on short-horizon tool use. Others emphasize red-teaming yield, classification quality, or refusal behavior. Because these metrics are usually reported without a stable lifecycle-wide evaluation framework, the raw amount of evaluation work currently exceeds the coherence of its conclusions. \tabref{tab:rq4-metrics} makes this imbalance visible by contrasting what the most common metrics capture well with what they systematically omit in practice.

\begin{table*}[!htbp]
\caption{Common Evaluation Metrics and Their Coverage Limits}
\scriptsize
\label{tab:rq4-metrics}
\centering
\setlength{\tabcolsep}{4pt}
\renewcommand{\arraystretch}{1.0}
\begin{tabularx}{\textwidth}{L{0.15\textwidth}C{0.04\textwidth}Y Y}
\toprule
\textbf{Metric} & \textbf{Count} & \textbf{What It Captures Well} & \textbf{What It Often Misses} \\
\midrule
Attack Success Rate & 129 & Whether an attack achieves its immediate objective & Utility loss, delayed harms, and deployment cost \\
\addlinespace[3pt]
Success Rate & 59 & Task completion under different settings & Whether the completed task remains aligned \\
\addlinespace[3pt]
Accuracy & 41 & Correctness on labeled tasks or benchmark items & Action risk and system-level consequences \\
\addlinespace[3pt]
Utility & 22 & Practical usefulness under defensive constraints & Fine-grained safety failure modes \\
\addlinespace[3pt]
Recall & 19 & Coverage of risky cases in detection tasks & Precision and downstream operational tradeoffs \\
\addlinespace[3pt]
F1 & 15 & Balance of precision and recall in classifications & Latency, cost, and task-level utility \\
\addlinespace[3pt]
Precision & 14 & False-alarm sensitivity in detectors and filters & Missed attacks and long-horizon effects \\
\addlinespace[3pt]
Refusal Rate & 13 & Degree of conservativeness under safety controls & Whether refusals are appropriate or overly broad \\
\addlinespace[3pt]
False Positive Rate & 13 & Unnecessary blocking or incorrect alarms & Broader task success and delayed compromise \\
\addlinespace[3pt]
Latency & 8 & Runtime overhead of a defense or monitor & Security quality and long-term utility \\
\addlinespace[3pt]
Cost & 7 & Resource burden of additional controls & Whether the expense improves real protection \\
\bottomrule
\end{tabularx}
\end{table*}

\subsection{Towards Stronger Evaluation}\label{sec:rq4-3}

The coverage and metric gaps identified above are not merely methodological refinements. They reveal a deeper conceptual gap because the field still lacks consensus on what secure agent behavior should look like once agents become stateful, tool-using, and collaborative. Benchmark fragmentation is therefore both a symptom and a cause of conceptual immaturity. It reflects the absence of a common evaluation theory for the agentic loop.
The benchmark coverage matrix confirms that reuse is densest around user- or web-facing inputs, planning, and tool execution, while memory-centric, inter-agent, coordination-centric, and deployment-cost dimensions remain sparse. \appref{sec:app-rq4-matrix} reports the full matrix, co-occurrence convention, and detailed interpretation.

This pattern also changes how benchmark scores should be interpreted. A single attack success rate can show that a defense fails somewhere, but often does not reveal where the failure entered the lifecycle, which authority boundary was crossed, whether the agent recovered, or what useful behavior was lost. Stronger evaluation should therefore treat a benchmark not only as a leaderboard task, but also as an instrumented diagnostic setting. Evaluations need step-level traces, explicit privilege contexts, recovery outcomes, and counterfactual utility after controls are applied. This is especially important for stateful and multi-agent settings, where a harmful effect may appear several turns after the triggering input and may be amplified by memory reuse, summarization, delegation, or shared communication channels. Without this diagnostic layer, the field risks optimizing for visible short-horizon failures while leaving unanswered whether an agent can preserve useful autonomy while containing contaminated information and unsafe authority transfer.

The same evidence points to a need for version-aware regression evaluation. Agent behavior can change when a model, prompt, tool, policy, memory mechanism, or orchestration component is updated, even when the intended task remains unchanged. The attack traces, poisoned states, privilege-sensitive actions, and multi-agent failures represented in current benchmarks can seed repeatable regression suites across such changes. Their value depends on preserving execution context and step-level evidence so later runs reveal not only score changes but whether previously protected lifecycle transitions have become vulnerable.

\subsection{Answer to RQ4}\label{sec:rq4-answer}

The evaluation ecosystem is active but not sufficiently standardized for deployment assurance. Stronger evaluation should meet five conditions. It should cover multiple threat surfaces rather than only prompt-level compromise. It should jointly report safety and utility. It should model longer horizons so that memory corruption and delayed activation become visible. It should include privilege-sensitive scenarios where external effects matter. It should also report latency, cost, and oversight burden consistently across studies.

\section{Discussion}\label{sec:discussion}

LLM agent security remains organized more around immediate problems than stable engineering principles. The field is increasingly effective at demonstrating manipulation, but less consolidated around architectures that encode authority, provenance, containment, and fallback behavior. Its central challenge is therefore to define well-governed agent behavior under realistic permissions, persistent state, and repeated deployment.

\subsection{Positioning Relative to Prior Surveys}\label{sec:discussion-1}

The corpus contains 32 survey papers, but lifecycle-based modeling, explicit annotation rules, benchmark synthesis, defense tradeoff analysis, and deployment assurance remain underdeveloped as a combined perspective. \tabref{tab:survey-positioning} compares representative broad agent-security surveys, LLM-agent security reviews, and multi-agent risk-management surveys across these dimensions. \appref{sec:app-survey-positioning} defines the qualitative rating criteria and row-assignment procedure.

The comparison reveals uneven specialization rather than an absence of prior coverage. Several surveys provide broad treatment of tools and runtime risks \cite{deng2025ai, he2026emerged, yu2025survey, zhu2025survey}. Among the representative set, \citet{zhu2025survey} gives particularly broad attention to memory and state, while \citet{raza2026trism} foregrounds multi-agent risk management. These strengths are complementary, but they are usually developed within topic-centered taxonomies rather than a shared model that traces how information, authority, and state move across the agentic loop.
This separation matters because coverage alone does not create cumulative evidence. Threats may be cataloged without identifying the lifecycle transitions they exploit, defenses may be grouped without making their trust assumptions or intervention points comparable, and benchmarks may be listed without showing which surfaces, stages, or deployment constraints they exercise. The consistently limited benchmark-synthesis column in \tabref{tab:survey-positioning} is therefore consequential. Without connecting security claims to evaluation coverage, it remains difficult to determine whether a proposed defense addresses the dominant attack paths or whether benchmark results support broader deployment claims.

\begin{table*}[!htbp]
\caption{Positioning This Survey Against Representative Prior Surveys}
\scriptsize
\label{tab:survey-positioning}
\centering
\setlength{\tabcolsep}{3pt}
\renewcommand{\arraystretch}{1.0}
\begin{tabularx}{\textwidth}{L{0.10\textwidth}Y Y Y Y Y Y Y}
\toprule
\textbf{Survey} & \makecell[l]{\textbf{Search}\\\textbf{Protocol}} & \makecell[l]{\textbf{Lifecycle}\\\textbf{Framing}} & \makecell[l]{\textbf{Tools /}\\\textbf{Runtime}} & \makecell[l]{\textbf{Memory /}\\\textbf{State}} & \makecell[l]{\textbf{Multi-}\\\textbf{agent}} & \makecell[l]{\textbf{Benchmark}\\\textbf{Synthesis}} & \makecell[l]{\textbf{Systems /}\\\textbf{Assurance}\\\textbf{Framing}} \\
\midrule
\cite{deng2025ai} & Partial & Partial & Broad & Partial & Partial & Limited & Limited \\
\addlinespace[3pt]
\cite{he2025security} & Limited & Limited & Partial & Limited & Limited & No & Limited \\
\addlinespace[3pt]
\cite{he2026emerged} & Partial & Partial & Broad & Limited & Partial & Limited & Limited \\
\addlinespace[3pt]
\cite{ji2024navigating} & Partial & Partial & Partial & Limited & Partial & No & Partial \\
\addlinespace[3pt]
\cite{yu2025survey} & Partial & Partial & Broad & Partial & Partial & Limited & Partial \\
\addlinespace[3pt]
\cite{liu2025comprehensive} & Partial & No & Partial & Partial & Partial & Limited & Limited \\
\addlinespace[3pt]
\cite{zhu2025survey} & Partial & Partial & Broad & Broad & Partial & Limited & Limited \\
\addlinespace[3pt]
\cite{raza2026trism} & Partial & Partial & Partial & Partial & Broad & Limited & Partial \\
\addlinespace[3pt]
\textbf{This survey} & \textbf{Explicit} & \textbf{Central} & \textbf{Broad} & \textbf{Broad} & \textbf{Broad} & \textbf{Dedicated} & \textbf{Dedicated} \\
\bottomrule
\end{tabularx}
\end{table*}

The contribution of this survey is thus relational rather than simply additive. Its lifecycle framing and corpus-level annotations connect threat surfaces to propagation paths, defenses to protected transitions and tradeoffs, and benchmarks to the risks they measure or omit. The systems and assurance perspective then carries these links into questions of privilege, persistent state, observability, and deployment evidence. The ratings in \tabref{tab:survey-positioning} characterize alignment with this analytical purpose rather than overall survey quality, and the earlier reviews remain valuable for the specialized areas they examine in greater depth.

\subsection{From Attack-Centric Analysis to Secure Agent Engineering}\label{sec:discussion-2}

The corpus suggests three design priorities for moving from attack-centric analysis to secure agent engineering. First, future systems need provenance-aware state management. Persistent memory, summaries, and scratchpads can preserve useful context, but they can also preserve contamination. Stronger mechanisms are needed for recording where a state item came from, under what authority it should be trusted, and when it should be revoked, decayed, or quarantined. The deeper issue is that memory changes the time scale of security. A single unsafe instruction may stop being an input event and become part of the agent's future operating context. Without a state model that carries provenance, authority, and expiration, memory protection will remain reactive, and long-horizon agents will lack a principled answer to what they are allowed to remember and reuse.
Second, secure tool governance should become a first-class architectural concern. Tool use is where many of the highest-consequence risks materialize, because it is also where language-mediated decisions cross capability boundaries. This suggests that agent frameworks should expose permissions, confirmation policies, capability scoping, and execution isolation as first-class design elements rather than as optional add-ons. The key research challenge is no longer simply whether a model can choose the right tool. It is whether the surrounding system can represent which authority is being delegated, which evidence justifies the delegation, and which effects must remain reversible or containable. The prominence of access control, information-flow control, and sandboxing in the corpus supports this direction, but current implementations remain uneven and often difficult to adapt to open-ended workflows.
Third, multi-agent systems require secure orchestration rather than coordination logic alone. Once tasks are delegated across agents, message integrity, role separation, authority scoping, and coordination topology become security concerns in their own right. Future work should therefore treat orchestration not only as a capability problem, but also as a trust-management problem. In this sense, multi-agent security is not just single-agent security repeated several times. It introduces new semantics of propagation, amplification, and failure coupling, where a local compromise can become a workflow-level failure because other agents inherit, reinterpret, or operationalize the compromised state.

\tabref{tab:discussion-agenda} condenses these priorities into a forward-looking research agenda. The common thread across the table is that the field's most urgent open problems all arise where state, authority, and evaluation interact. Progress will depend less on inventing another isolated attack or defense category, and more on building a security control plane for agents. Such a control plane would make provenance, permissions, memory lifecycle, communication boundaries, runtime policy, telemetry, and recovery visible to both developers and evaluators. This is the missing layer between model-level safety and deployment-level assurance.

\begin{table}[!htbp]
\caption{Open Challenges and Research Priorities for Secure LLM Agents}
\scriptsize
\label{tab:discussion-agenda}
\centering
\setlength{\tabcolsep}{4pt}
\begin{tabularx}{\columnwidth}{L{0.2\columnwidth}Y Y}
\toprule
\textbf{Open Challenge} & \textbf{Why Current Evidence Is Insufficient} & \textbf{Promising Direction} \\
\midrule
Provenance-aware state management & Existing work treats stored context as useful history more often than as security-relevant state & Memory provenance, authority tags, trust decay, revocation, and replayable state histories \\
\addlinespace[3pt]
Secure tool governance & Many studies expose unsafe tool use without specifying how authority should be delegated, constrained, or recovered & Fine-grained capability scoping, effect typing, confirmation policies, and reversible execution paths \\
\addlinespace[3pt]
Multi-agent trust management & Current work highlights propagation and role confusion, but lacks models of how trust changes across delegation chains & Role-scoped authority, message integrity controls, trust attenuation, and topology-aware containment \\
\addlinespace[3pt]
Lifecycle-wide evaluation & Benchmarks often observe whether an attack succeeds, but less often explain which lifecycle transition failed & Longer-horizon suites with step-level traces, memory reuse, coordination, and privilege-sensitive scenarios \\
\addlinespace[3pt]
Deployment-oriented assurance & Current evaluations rarely connect benchmark outcomes to concrete deployment claims and operational responsibilities & Assurance cases that combine threat modeling, runtime controls, telemetry, fallback behavior, and empirical evidence \\
\bottomrule
\end{tabularx}
\end{table}

\subsection{Implications for Agent Systems Research and Practice}\label{sec:discussion-3}

For researchers and practitioners working on agent systems, the value of this survey lies not only in cataloging new attacks, but also in translating them into recurring obligations for architecture, implementation, evaluation, deployment, and system evolution.

The first implication concerns requirements and architecture. Agent systems need security requirements that explicitly bind \emph{authority}, \emph{state}, and \emph{tool scope}. Traditional functional requirements such as ``the agent can browse the web'' or ``the agent can edit code'' are too coarse for secure deployment. They should be refined into authority-scoped requirements, \eg which tools may be called under which provenance conditions, which memory items may be reused across tasks, which actions require confirmation, and what rollback or containment behavior should occur after a suspected compromise. This suggests a clear research opportunity for agent-oriented threat modeling and secure agent API design.

The second implication concerns development workflows, especially coding agents and CI/CD integration. The corpus shows that software-engineering scenarios are both high-frequency and high-consequence. Coding agents often combine file access, shell execution, package installation, secret exposure, and tool-mediated prompt injection in the same loop. From a software engineering perspective, this means that secure-by-design coding agents should expose least-privilege execution, provenance-aware file access, typed tool contracts, and auditable approval boundaries as framework defaults rather than optional hardening layers. Supply-chain risk is part of the same picture, since MCP tools, skills, plugins, and repositories extend the attack surface in ways that resemble dependency risk in conventional software ecosystems, but with more ambiguity about what counts as data versus executable control.

The third implication concerns testing and regression assurance. Many current evaluations still resemble one-off red-teaming studies, whereas engineering teams need repeatable regression tests and sustainable maintenance practices. For agentic systems, this points towards security regression suites that replay prompt-injection traces, malicious tool outputs, poisoned memory states, and multi-agent communication failures across versions of prompts, tools, policies, and orchestration logic. Benchmarks such as \textit{AgentDojo}, \textit{ASB}, \textit{OS-Harm}, and \textit{AgentLeak} are useful seeds, but they need to be connected to CI-style security gates, step-level telemetry, and failure triage workflows before they can support mature software processes.

The fourth implication concerns runtime verification and operations. Because many agent failures are stateful and delayed, verification cannot stop at pre-deployment testing. Runtime monitors, policy engines, and containment layers must become first-class operational components. This connects established work on runtime verification, self-adaptive systems, and assurance cases with broader concerns in artificial intelligence, systems security, and software engineering. In particular, long-horizon agent deployment calls for provenance-aware memory management, explicit tool-approval policies, post-incident replayability, and maintenance procedures for updating prompts, tools, and policies without silently widening the authority surface.

Taken together, these implications suggest that LLM agent security should be treated as a cross-lifecycle systems problem involving requirements engineering for trust boundaries, secure design for tool and memory mediation, regression testing for long-horizon use, runtime verification for consequential execution, and maintenance for evolving agent ecosystems in real deployments.

\subsection{Research Priorities for the Next Stage}\label{sec:discussion-4}

The evaluation and engineering concerns discussed above point to broader research priorities for high-impact domains. Software engineering, finance, healthcare, and embodied robotics all require stronger evidence than short-horizon toy benchmarks can provide. In such settings, the relevant question is not only whether one attack succeeds, but whether the system can remain governable under imperfect information, partial compromise, repeated use, and nontrivial operational constraints. This changes the unit of progress. The next stage should not be measured only by larger benchmark sets or lower attack-success rates, but by whether the field can explain why a control works, where it fails, how it composes with other controls, and what evidence justifies deployment.

For this reason, a natural next step is the development of deployment-oriented assurance cases for LLM agents. An assurance case would combine threat modeling, privilege analysis, runtime controls, benchmark evidence, telemetry requirements, and incident-containment expectations for a specific deployment setting. Such a direction is well aligned with software engineering and systems research because it connects design choices to operational evidence rather than treating evaluation as an isolated leaderboard exercise. It also offers a way to bridge the gap between broad survey taxonomies and deployment-specific claims by making explicit what assumptions, controls, and evidence are required before an agent should be trusted in a given environment.

A second priority is compositional evaluation. Many defenses look promising when tested against one attack surface, one benchmark, or one agent architecture, but deployment risk emerges from interactions among memory, tools, delegation, and oversight. Future evaluations should therefore ask whether controls preserve their meaning when combined. A permission boundary may depend on provenance. A memory filter may depend on tool-output labeling. A multi-agent containment policy may depend on whether messages carry authority metadata. These dependencies suggest that benchmark design should move from isolated scenarios towards stress tests of security invariants across the whole agentic loop.

A third priority is studying recovery, not only prevention. Real deployments will face ambiguous alerts, partial failures, and uncertain compromise boundaries. Yet much of the current literature stops at attack success, defense success, or task utility. Mature agent security will require evidence about how systems detect suspicious state, roll back unsafe actions, quarantine contaminated memory, reconstruct decision traces, and continue operation under reduced authority. This recovery perspective is especially important for long-running agents because the cost of failure depends not only on whether compromise occurs, but also on whether its effects can be bounded after discovery.

Overall, the discussion points to a consistent theme. The field will mature when secure LLM agents are treated as engineered systems with explicit trust boundaries, governable state, constrained authority, observable execution, and deployment-specific assurance arguments. The most important future work is therefore not a single stronger guardrail, but a shift in abstraction. LLM agent security needs concepts and tools that make language-mediated authority as explicit, testable, and maintainable as other security-critical interfaces in software systems.

\section{Threats to Validity}\label{sec:validity}

\textbf{Construct validity.} The most important construct-validity risk is category overlap. Threat surfaces such as web content, retrieved content, tool outputs, files, and memory can all appear as untrusted text, while lifecycle stages such as planning, decision, and tool execution are often tightly interleaved in modern agent frameworks. We mitigate this risk by publishing explicit operational rules, using multi-label annotation for technical fields, and separating descriptive single-label fields from substantive multi-label annotations. Even so, some boundary cases remain judgment-sensitive.

\textbf{Internal validity.} The corpus relies on manual screening and manual annotation. Relevance judgments, label assignment, and later harmonization therefore involve human interpretation.
To reduce subjectivity, all 247 papers were independently annotated by three authors using conservative inclusion rules, explicit exclusion categories, version normalization rules, and a boundary-aware annotation guide.
Agreement on the main annotation fields reached a Fleiss' $\kappa$ of 0.86 before disagreement resolution, and remaining cases were harmonized through discussion.

\textbf{External validity.} Search bias and publication bias are unavoidable in a fast-moving area. The search protocol spans six major sources and includes snowballing and LLM-assisted candidate expansion, but it may still miss papers that use unusual terminology, unpublished industrial artifacts, or adjacent system names not captured by the query template. The field is also heavily preprint-based, with 169 of 247 papers from arXiv. This improves timeliness but weakens stability because titles, claims, and experimental designs may change before archival publication.

\textbf{LLM-assisted retrieval bias.} The candidate-expansion stage uses GPT-5.4 with web search to surface lexically distant papers and query variants. This improves recall, but it also introduces model bias because the model may over-suggest highly visible benchmarks, popular terminology, or papers that are easy to retrieve on the public web. We therefore limit the LLM's role to candidate suggestion and term normalization and require manual verification for every candidate that enters the screening pool for review.

\section{Conclusion}\label{sec:conclusion}

LLM agent security has emerged as a distinct research area because LLM agents transform language into decisions, tool invocations, persistent state, and coordinated action. Based on a corpus of 247 papers, this survey argues that the field is best understood through a lifecycle-based and surface-aware framework that connects information flow, delegated authority, and persistent state across the agentic loop.
Viewed through that framework, three descriptive conclusions are especially robust. First, prompt injection and tool-mediated control-flow hijacking still define the empirical center of the reviewed threat literature. Second, the clearest expansion visible in the corpus lies in persistence and propagation, especially through memory reuse and multi-agent coordination. Third, the current defense and evaluation ecosystems are less consolidated than the threat literature they respond to, since defense research offers useful components but not yet a stable compositional stack, and evaluation research offers substantial activity but not yet a common basis for deployment assurance.
Secure LLM agents will therefore require more than better prompts or stronger filters. They will require principled privilege control, provenance-aware state management, runtime oversight, and evaluation practices that reflect realistic operational conditions.
More broadly, they will require a shift from securing model outputs in isolation to engineering governable agentic systems whose authority, memory, and coordination structure are treated as first-class concerns in system design, evaluation, deployment, and evolution.

\begin{acks}
This work is partially supported by the National Key Research and Development Program of China (2024YFF0908000).
\end{acks}

\bibliographystyle{ACM-Reference-Format}
\bibliography{main}

\clearpage
\appendix

\section*{APPENDIX}

\section{Review Protocol Details}\label{sec:app-method}

\subsection{Search Query and LLM-Assisted Expansion}\label{sec:app-search}

To improve recall, the canonical Boolean template combines LLM, agent, and security concepts.
\begin{querybox}
\small
(\texttt{"large language model"} OR \texttt{"language model"} OR \texttt{LLM}) AND \\
(\texttt{agent} OR \texttt{"tool-using agent"} OR \texttt{"web agent"} OR \texttt{"coding agent"} OR \texttt{"browser agent"} OR \texttt{"computer-use agent"} OR \texttt{"memory-augmented agent"} OR \texttt{"embodied agent"} OR \texttt{"multi-agent"} OR \texttt{"agentic workflow"}) AND \\
(\texttt{security} OR \texttt{safety} OR \texttt{"prompt injection"} OR \texttt{jailbreak} OR \texttt{"memory poisoning"} OR \texttt{"tool misuse"} OR \texttt{"data exfiltration"} OR \texttt{"access control"} OR \texttt{sandboxing} OR \texttt{"runtime monitoring"} OR \texttt{guardrail} OR \texttt{benchmark} OR \texttt{"red teaming"})
\end{querybox}
The same normalized term groups are retained when searches are re-run after terminology expansion. Benchmark names, framework names, and application-specific terms serve as discovery cues without becoming eligibility criteria.

The retrieval assistant receives the current seed list together with titles, keywords, publication dates, and the three concept groups. It is asked to find lexically distant but semantically related work, identify benchmark aliases, normalize synonymous terms, and flag missing branches such as memory poisoning, instruction hierarchy, MCP tooling, and multi-agent governance. Suggested terms are used to re-run targeted searches, while suggested papers enter screening only after their bibliographic identity and relevance have been checked against source records.

\begin{querybox}
\small
You are assisting a literature review on LLM agent security. Given the current seed papers, the search window, and the concept groups for LLMs, agents, and security, suggest additional candidate papers and search terms that may be missed by exact keyword retrieval. Focus on papers about tool-using agents, web agents, coding agents, memory-augmented agents, embodied agents, MCP or plugin ecosystems, and multi-agent workflows. Include work on attacks, defenses, benchmarks, evaluations, and system or deployment reports. For each candidate, provide the title, year, venue or preprint source, a short reason for relevance, and the query term or known benchmark name that led to it. Do not decide inclusion. Mark uncertain candidates as uncertain so that they can be checked manually.
\end{querybox}

Backward snowballing traces references that establish attack, defense, and benchmark lineages. Forward snowballing identifies later work that reuses or extends those foundations. Both directions recover influence expressed through system lineage, benchmark reuse, or architectural development rather than literal query overlap.

\subsection{Exclusion Breakdown and Selection Boundary}\label{sec:app-selection}

\begin{table}[!htbp]
\caption{Exclusion Decisions by Principle}
\scriptsize
\label{tab:selection-detail}
\centering
\setlength{\tabcolsep}{3pt}
\begin{tabular}{lc}
\toprule
\textbf{Exclusion Principle} & \textbf{Records} \\
\midrule
Generic prompt-injection or instruction-separation work without an explicit agentic loop & 12 \\
Broad LLM safety, privacy, governance, or position papers with indirect relevance & 5 \\
Agentic or autonomous-system papers where security was not central & 4 \\
Benchmarks primarily measuring capability rather than safety or security & 3 \\
\bottomrule
\end{tabular}
\end{table}

Each excluded record is assigned to one principle, so the four rows partition the 24 removals. Adjacent safety, governance, and capability literatures remain informative, but they do not all support the agentic-loop annotation frame used in this review. Search-engine hit counts are not treated as stable evidence because they vary with date, interface, personalization, and query syntax.

\subsection{Annotation Encoding, Boundary Cases, and Reliability}\label{sec:app-annotation}

Field-level agreement is calculated before harmonization. Single-label fields use one categorical value per paper, while each controlled label in a multi-label field is encoded as a separate present-or-absent decision. The resolution focus in \tabref{tab:annotation-reliability} records the boundary cases discussed before the final label set was fixed.
The three independent annotations were compared before adjudication. Remaining disagreements were discussed jointly and harmonized only after the authors reached a shared interpretation of the relevant boundary rule.

The field-level pattern reflects semantic granularity. Agreement is highest for system setting and primary paper type, which require one principal assignment. Multi-label fields demand finer judgments about compound contributions and neighboring labels, explaining variation among lifecycle, threat, attack, and defense fields. The values should therefore be interpreted by field rather than as a uniform level of reliability. All reported $\kappa$ values describe independent pre-harmonization decisions, whereas the synthesis uses adjudicated labels.

\begin{table}[!htbp]
\caption{Annotation Reliability by Field}
\scriptsize
\label{tab:annotation-reliability}
\centering
\setlength{\tabcolsep}{3pt}
\begin{tabularx}{\columnwidth}{L{0.2\columnwidth}L{0.2\columnwidth}C{0.1\columnwidth}Y}
\toprule
\textbf{Field} & \textbf{Encoding for agreement} & \textbf{Fleiss' $\kappa$} & \textbf{Resolution focus} \\
\midrule
Primary paper type & Single categorical label & 0.90 & Boundary cases among attack, defense, benchmark, evaluation, report, and survey papers \\
System setting & Single categorical label & 0.96 & Single-agent versus multi-agent focus \\\addlinespace[3pt]
Lifecycle stage & Multi-label binary vector & 0.84 & Overlap among planning, decision, tool execution, memory, monitoring, and coordination \\
Threat surface & Multi-label binary vector & 0.81 & Provenance differences among prompts, web content, tool outputs, files, memory, and inter-agent channels \\
Attack family & Multi-label binary vector & 0.80 & Co-occurring injection, malicious tools, poisoning, exfiltration, and coordination failures \\
Defense method & Multi-label binary vector & 0.78 & Compound defenses combining guardrails, access control, sandboxing, monitoring, and provenance \\
Benchmark usage & Multi-label binary vector & 0.87 & Papers using multiple benchmarks or custom variants \\\addlinespace[3pt]
Evaluation metric & Multi-label binary vector & 0.82 & Distinguishing safety, utility, attack success, latency, cost, and oversight measures \\
\bottomrule
\end{tabularx}
\end{table}

\subsection{Rating Rubric for Prior-Survey Positioning}\label{sec:app-survey-positioning}

For methodological dimensions, \textit{Explicit} means that a survey presents a clearly identifiable procedural or analytical treatment of the dimension, such as an articulated search protocol, annotation guide, or lifecycle framework. \textit{Central} means that the dimension is not only explicit but also functions as a primary organizing backbone across the survey. \textit{Partial} means that the dimension is discussed but not as a stable organizing device or not with enough procedural detail to support audit. \textit{No} means that the dimension is absent as an identifiable analytical element. For scope dimensions, \textit{Broad} means that the survey substantially covers multiple agent forms or multiple security-relevant subtopics under that dimension, \textit{Partial} means visible but bounded treatment, and \textit{Limited} means only brief or peripheral coverage. For synthesis dimensions, \textit{Dedicated} means that the survey gives the dimension a recurring analytical role or a standalone synthesis focus rather than mentioning it only in passing. Row-level ratings were assigned by checking whether each survey devotes sustained subsection-, table-, or protocol-level treatment to the dimension. The ratings remain qualitative, but they are anchored in explicit criteria rather than impressionistic labels.

Application is dimension-specific. Search Protocol assesses visible retrieval, screening, and analysis, while Lifecycle Framing asks whether stages organize the synthesis. Scope columns reflect sustained breadth. Benchmark Synthesis requires cross-resource comparison, and Systems and Assurance Framing requires recurring trust boundaries, controls, or deployment evidence. Ratings characterize analytical emphasis rather than overall survey quality.

\section{Normalized Venue Distribution}\label{sec:app-trend}

\begin{table}[H]
\caption{Venue Distribution of Reviewed LLM Agent Security Papers by Publication Type}
\label{tab:app-trend}
\centering
\resizebox{\textwidth}{!}{
\scriptsize
\renewcommand{\arraystretch}{1.2}
\begin{tabular}{p{0.16\textwidth}p{0.7\textwidth}p{0.11\textwidth}}
\toprule
\textbf{Venue Abbr.} & \textbf{Full Venue or Source Name} & \textbf{Publications} \\\midrule
\multicolumn{3}{l}{\textit{Informal Publications}} \\
\addlinespace[3pt]
arXiv & arXiv preprint & 169 (68.42\%) \\
Web & Web-published industry reports, blogs, and advisories & 12 (4.86\%) \\
\textbf{Subtotal} & & \textbf{181 (73.28\%)} \\\midrule
\multicolumn{3}{l}{\textit{Conference Papers}} \\
\addlinespace[3pt]
ICLR & International Conference on Learning Representations & 12 (4.86\%) \\
ACL & Annual Meeting of the Association for Computational Linguistics & 10 (4.05\%) \\
EMNLP & Conference on Empirical Methods in Natural Language Processing & 8 (3.24\%) \\
NeurIPS & Conference on Neural Information Processing Systems & 6 (2.43\%) \\
ICML & International Conference on Machine Learning & 5 (2.02\%) \\
NDSS & Network and Distributed System Security Symposium & 3 (1.21\%) \\
COLM & Conference on Language Modeling & 2 (0.81\%) \\
NAACL & Annual Conference of the North American Chapter of the Association for Computational Linguistics & 2 (0.81\%) \\
AAAI & AAAI Conference on Artificial Intelligence & 1 (0.40\%) \\
ACM MM & ACM International Conference on Multimedia & 1 (0.40\%) \\
AIES & AAAI/ACM Conference on AI, Ethics, and Society & 1 (0.40\%) \\
ICSE & International Conference on Software Engineering & 1 (0.40\%) \\
IJCAI & International Joint Conference on Artificial Intelligence & 1 (0.40\%) \\
ICRA & IEEE International Conference on Robotics and Automation & 1 (0.40\%) \\
ACM CCS & ACM Conference on Computer and Communications Security & 1 (0.40\%) \\
\textbf{Subtotal} & & \textbf{55 (22.27\%)} \\\midrule
\multicolumn{3}{l}{\textit{Journal Papers}} \\
\addlinespace[3pt]
ICT Express & ICT Express & 1 (0.40\%) \\
AI & AI, MDPI journal & 1 (0.40\%) \\
Information Fusion & Information Fusion & 1 (0.40\%) \\
Information Sciences & Information Sciences & 1 (0.40\%) \\
ACM CSUR & ACM Computing Surveys & 1 (0.40\%) \\
HCC & High-Confidence Computing & 1 (0.40\%) \\
FCS & Frontiers in Computer Science & 1 (0.40\%) \\
JSS & Journal of Systems and Software & 1 (0.40\%) \\
FGCS & Future Generation Computer Systems & 1 (0.40\%) \\
JAIP & Journal of Artificial Intelligence Practice & 1 (0.40\%) \\
FnT P\&S & Foundations and Trends in Privacy and Security & 1 (0.40\%) \\
\textbf{Subtotal} & & \textbf{11 (4.45\%)} \\\midrule
\textbf{Total} & & \textbf{247 (100.00\%)} \\
\bottomrule
\end{tabular}}
\end{table}

Findings papers are grouped with their parent ACL-family venue. Workshop papers are grouped with the corresponding parent venue only when that relationship is explicit. Web-published system reports, blogs, and advisories are grouped as Web. Duplicate versions and preprint-publication pairs are treated as one bibliographic identity during final normalization. Each retained study therefore contributes once to the venue table under its final normalized identity, and the publication-type subtotals partition the 247-paper corpus without counting preprint and archival versions separately.

Venue normalization remains separate from paper-type annotation, which records research function rather than dissemination source. Outside arXiv and Web, 66 papers appear across 26 formal venues. ICLR, ACL, EMNLP, NeurIPS, and ICML contribute 41 of the 55 conference papers, the other ten conference venues contribute 14, and the 11 journal papers occur one per outlet. Naming every outlet avoids an Others category, separates functional coding from publication identity, and makes the totals directly auditable.

\section{Supplementary Evidence for the Research Questions}\label{sec:app-rq-evidence}

\subsection{Lifecycle-by-Surface Matrix}\label{sec:app-rq1-matrix}

\begin{figure}[H]
\centering
\includegraphics[width=0.94\linewidth]{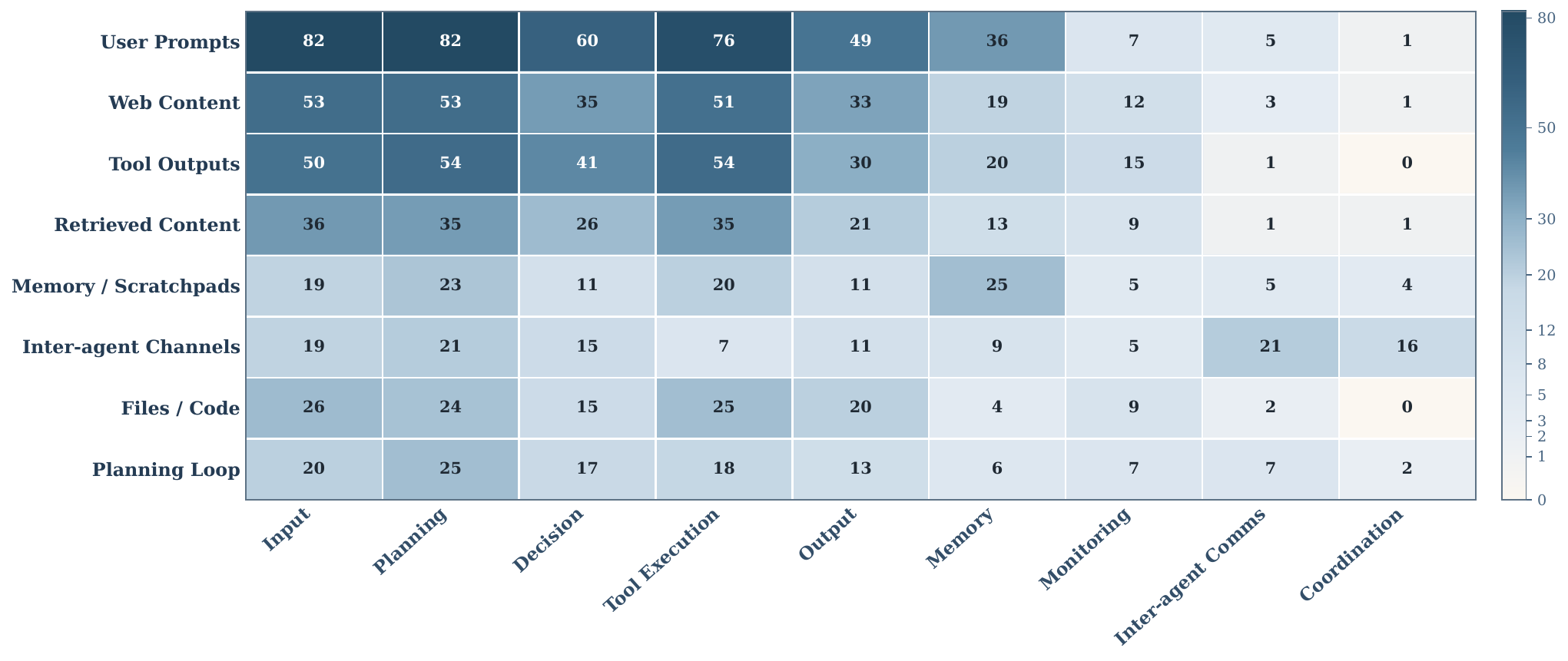}
\vspace{-3mm}
\caption{Lifecycle-by-Surface Matrix of Major Threat Surfaces and Focal Lifecycle Stages}
\label{fig:app-rq1-model}
\end{figure}

Each cell is a paper-level co-occurrence count between one controlled threat-surface label and one lifecycle-stage label. A paper may contribute to several cells because both fields are multi-label. Row or column sums therefore describe research attention across coded combinations rather than mutually exclusive partitions of the corpus.

The matrix shows that \textit{User Prompts} concentrate heavily in \textit{Input} and \textit{Planning} with 82 papers in each cell, but remain substantial in \textit{Decision} with 60, \textit{Tool Execution} with 76, and \textit{Output} with 49. Direct prompting is therefore often a cross-stage control problem rather than a single entry event. \textit{Web Content} and \textit{Tool Outputs} also concentrate in \textit{Input}, \textit{Planning}, and \textit{Tool Execution}, making visible the mediated paths through which externally sourced text shapes agent behavior. \textit{Memory / Scratchpads} align most strongly with \textit{Memory}, while \textit{Inter-agent Channels} align most strongly with \textit{Coordination}. These combinations distinguish immediate entry paths from risks that depend on persistence or propagation.

\subsection{Scenario-Oriented Defense Stacks}\label{sec:app-rq3-scenarios}

\begin{table}[H]
\caption{Scenario-Oriented Minimal Defense Stacks by Agent Class}
\fontsize{6.7}{7.1}\selectfont
\label{tab:rq3-scenario-stacks}
\centering
\setlength{\tabcolsep}{4pt}
\renewcommand{\arraystretch}{0.9}
\begin{tabularx}{\textwidth}{L{0.10\textwidth}L{0.21\textwidth}Y L{0.30\textwidth}}
\toprule
\textbf{Agent Class} & \textbf{Dominant Tuple Exposure} & \textbf{Likely Necessary Layers} & \textbf{Main Open Problems} \\
\midrule
Browser and web agents & $I \rightarrow P \rightarrow D \rightarrow T \rightarrow O$ with untrusted web content & Input-trust management, runtime plan and action checks, tool gating or confirmation, execution containment for high-authority actions & Robust source separation under realistic browsing, high-utility defenses against indirect prompt injection \\
\addlinespace[3pt]
Coding agents & $I \rightarrow P \rightarrow D \rightarrow T \rightarrow O$ plus files, code, and local execution & Least privilege for file, shell, and package operations, runtime monitoring of dangerous tool calls, and isolated execution for code and commands & Secure-by-default tool APIs, dependency and skill provenance, regression testing for long tool chains \\
\addlinespace[3pt]
Memory-augmented assistants & $I \leftrightarrow M \rightarrow P \rightarrow D \rightarrow T$ with persistent state reuse & Provenance-aware state isolation, revocation or decay mechanisms, runtime checks before memory-derived actions & Trust labeling for stored state, delayed-trigger evaluation, memory-specific benchmark realism \\
\addlinespace[3pt]
Multi-agent workflows & $C \rightarrow P/D/T$, $M \leftrightarrow C$, cross-agent propagation & Role-scoped authority, communication monitoring, topology-aware containment, per-agent privilege control & Covert collusion, message provenance, distributed rollback, incident containment across agent graphs \\
\bottomrule
\end{tabularx}
\end{table}

The table is an engineering interpretation of the corpus rather than a validated standard. The recurring tradeoffs concern \textit{scope}, \textit{trust assumption}, \textit{overhead}, and \textit{utility impact}. Scope distinguishes protection of prompt interpretation, capability boundaries, state reuse, and execution environments. Trust assumptions distinguish model-mediated enforcement from policies, monitors, and isolated infrastructure. Stronger mediation generally adds latency or architectural complexity, while restrictive controls may reduce autonomy, flexibility, or task completion.

\subsection{Benchmark Coverage Matrix}\label{sec:app-rq4-matrix}

\begin{figure}[H]
\centering
\includegraphics[width=0.99\textwidth]{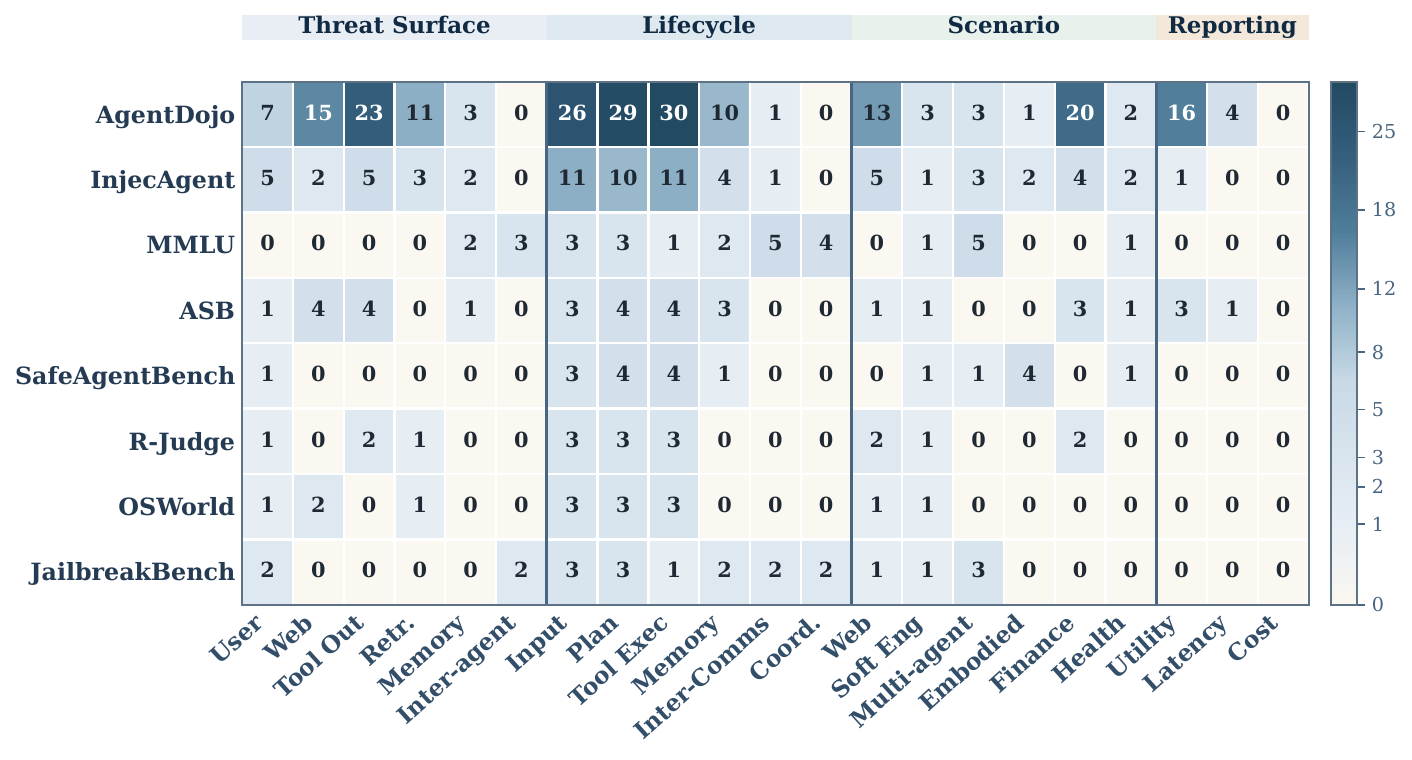}
\vspace{-2mm}
\caption{Benchmark Co-occurrence Counts Across Threat Surfaces, Lifecycle Stages, Scenarios, and Reporting Dimensions}
\label{fig:app-rq4-matrix}
\end{figure}

Each matrix cell reports the number of reviewed papers in which a named benchmark co-occurs with the corresponding multi-label code. One paper may contribute to several cells within a row and to several rows when it uses multiple benchmarks. The matrix consequently records corpus-level co-occurrence rather than an intrinsic property or quality score of a benchmark.

\textit{AgentDojo} and \textit{InjecAgent} co-occur frequently with user- or web-facing inputs, planning, and tool execution. Counts are much thinner for memory-centric, inter-agent, and coordination-centric settings. Reporting coverage is narrower still. Utility appears unevenly, latency appears rarely, and cost is nearly absent. Benchmark reuse is therefore most developed around the short-horizon pathways already emphasized by the literature and least developed where state, coordination, and deployment constraints require broader evidence.

The row profiles reveal further differences. \textit{AgentDojo} has the broadest pattern, co-occurring with tool outputs in 23 papers, input in 26, planning in 29, tool execution in 30, finance in 20, and utility in 16. It is also the only benchmark with latency reported in more than one paper. \textit{InjecAgent} covers input, planning, and tool execution in 11, 10, and 11 papers but has thinner reporting coverage. Among more specialized benchmarks, \textit{MMLU} is most visible in inter-agent communication, coordination, and multi-agent scenarios, while \textit{SafeAgentBench} has its strongest scenario count in embodied settings. Benchmark reuse therefore does not imply interchangeable coverage.

\end{document}